\documentclass{ws-jtca}

\usepackage{cases}
\usepackage{overpic}
\usepackage{subfigure} 
\usepackage{booktabs}
\usepackage{lineno}
 \usepackage{epstopdf}
\usepackage{colordvi}
\usepackage{color,xcolor}
\usepackage{amsmath}
\usepackage{bm}
\definecolor{re1}{RGB}{230,0,0}
\definecolor{re2}{RGB}{0,128,0}
\begin{document}

\markboth{}{Spectral Element Method for the Elastic/Acoustic Waveguide Problem in Anisotropic Metamaterials}

%
\catchline{}{}{}{}{}
%

\title{Spectral Element Method for the Elastic/Acoustic Waveguide Problem in Anisotropic Metamaterials}

\author{An Qi Ge}

\address{School of Mathematical Sciences, Xiamen University,\\
Institute of Electromagnetics and Acoustics, College of Electronic Science and Technology, Xiamen University\\
Xiamen, 361005, China\\
\email{angelge97@163.com} }

\author{Ming Wei Zhuang}
\address{Institute of Electromagnetics and Acoustics, College of Electronic Science and Technology, Xiamen University\\
Xiamen, 361005, China\\
\email{mw.zhuang@xmu.edu.cn} }

\author{Jie Liu *}
\address{Institute of Electromagnetics and Acoustics, College of Electronic Science and Technology, Xiamen University\\
The Postdoctoral Mobile Station of Information and Communication Engineering, School of Informatics,  Xiamen University\\
Xiamen, 361005, China\\
\email{liujie190484@163.com} }

\author{Qing Huo Liu \footnote{Corresponding authors}}
\address{Department of Electrical and Computer Engineering, Duke University\\
Durham NC, 27708, USA\\
\email{qhliu@duke.edu}}

\maketitle


\begin{abstract}
In order to simulate elastic wave propagation in a complex structure with inhomogeneous media, we often need to obtain the propagating eigenmodes of an elastic waveguide. As the waveguide is assumed uniform in one direction, the original 3-D problem can be converted into a so-called 2.5-D problem by using the Fourier transform in that direction. However, the introduction of elastic metamaterials (EMM) broadens the horizon of this subject, and new features are required in EMM waveguides that cannot be obtained by most traditional waveguide solvers. In this work, a spectral element method (SEM) is developed to simulate the elastic/acoustic waveguide problem in anisotropic media with anisotropic mass density and/or negative index parameters. To the best of our knowledge,  the SEM has not been introduced previously for such a waveguide problem. For waveguides with anisotropic density that cannot be solved by the FEM in most of commercial software packages, we design an anisotropic density EMM waveguide with our SEM solver to demonstrate some intriguing phenomena. The spectral element results are verified by several numerical examples through comparison with the traditional finite element method (FEM) to show its significant advantages in term of accuracy and computation efficiency.
\end{abstract}
\section{Introduction}

Recently, elastic waveguide problems have gained much attention due to various engineering applications, for example, ultrasound characterization, non-destructive testing, and structural health monitoring \cite{Baron2010Propagation,su2006guided,banerjee2009wave}. For various types of elastic waveguides \cite{MiklowitzThe}, the mode analysis is an important research topic, because wave propagation and scattering phenomena in a waveguide can be described as the superposition of all of the propagation modes and evanescent modes. In this class of problems, one is interested in solving the propagation constants and the corresponding field distributions of individual modes in a given waveguide structure. 

As an extension of the electromagnetic and acoustic waveguides \cite{LiuMixed,KirbyTransmission}, Lagasse proposes a finite element method (FEM) for computing the eigenmodes of the homogeneous elastic waveguides of arbitrary cross sections \cite{lagasse1973higher}; Kosmodamianskii {\it et. al.} derive the equation for normal elastic waves in a longitudinally
anisotropic cylindrical waveguide with a circular cross-section \cite{kosmodamianskii2005dispersion}; and Gravenkamp {\it et. al.} develop the scaled boundary finite element method (SBFEM) for an inhomogeneous isotropic elastic waveguide \cite{gravenkamp2013computation}. A semi-analytical finite element method (SAFEM) has been developed to simulate the solid-fluid coupling waveguide \cite{HLADKYHENNION1998265,MAZZOTTI2014408,VAZIRIASTANEH2017200} and an open waveguide with the absorbing boundary condition (ABC)  \cite{G2014}. Expressly, the spectral element method (SEM) is also used to solve the piezoelectric waveguide problem \cite{LIANG2019205} which can derive the elastic waveguide. All of these investigations focus on specific problems with significant applications.

Recently, with the advent of elastic metamaterials (EMM),  various new and intriguing wave propagation phenomena can be generated in such novel synthetic materials  \cite{LiuAn,ZhuA,ZhuMicrostructural,WuElastic}. As EMMs can involve negative refractive index materials and/or anisotropic mass density, most traditional numerical waveguide solvers such as the finite difference (FD) method \cite{SunPropagation} , FEM \cite{MoserModeling,TreysseNumerical,bartoli2006modeling} and SEM \cite{LIANG2019205,2012Guided} have not been modified to accommodate the modeling of such EMM waveguides.

In recent years, the SEM has been successfully developed for analyzing the propagation and scattering of elastic waves \cite{ShiSpectral,Komatitsch1999The,komatitsch2000wave,HU2020112761}, acoustic waves \cite{seriani1994spectral,seriani2008dft}, and electromagnetic waves \cite{1542240,LeeA}. As a special version of the high-order FEM, the SEM not only takes advantage of the geometric flexibility of the FEM, but also has the high accuracy of spectral methods. Moreover, the use of the Gauss-Lobatto-Legendre (GLL) polynomials for the nodal basis functions makes the relative errors converge exponentially with the order of basis functions.
Therefore, at the same accuracy requirement, the SEM requires much fewer degrees of freedom (DOFs) than the FEM, so it can save huge computational costs.

This investigation aims at developing an effective SEM for an inhomogeneous elastic waveguide of an arbitrary cross section applicable to both anisotropic solid materials and all kinds of EMMs, for instance, negative index, anisotropic mass density and so on, for the first time. In addition, the waveguide system with both solids and fluids is also treated by considering the fluid-solid coupling.  The external boundary of the waveguide cross section can allow different boundary conditions (BC) based on the given general formulations of the boundary terms, for example, the hard BC, the soft BC, the Bloch periodic boundary condition (BPBC) and the absorbing boundary condition (ABC) for an unbounded (open) waveguide. Numerical examples show the significant advantages of the SEM in term of accuracy and computation efficiency compared with the conventional FEM as implemented by COMSOL. We also design the simulation of waveguides with anisotropic density that cannot be solved by commercial solvers.

The organization of this paper is as follows. In Section II, we will present the detailed weak formulation of the elastic waveguide. The discretization by the SEM is shown in Section III. Finally, the accuracy and efficiency of the SEM are demonstrated by several examples in Section IV.

\section{GOVERNING EQUATIONS AND WEAK FORMULATIONS}
 \subsection{Governing Equations}
 For a general anisotropic and inhomogeneous elastic metamaterial with a potentially anisotropic mass density, elastic wave equations in frequency domain read
\begin{subnumcases}{}
	\omega^2\boldsymbol{\rho}\cdot\boldsymbol{u}+\boldsymbol{\nabla}\cdot\boldsymbol{\tau}={\bold 0}\label{wave equation}\\
	\boldsymbol{\tau}=\boldsymbol{c}:\boldsymbol{\epsilon}\label{tau}\\
	\boldsymbol{\epsilon}=(\boldsymbol{\nabla}\boldsymbol{u}+\boldsymbol{\nabla}\boldsymbol{u}^T)/2\label{epsilson}
 \end{subnumcases}
 where $\boldsymbol{\rho}=(\rho_{ij})_{3\times3}$ is the anisotropic mass density; $\omega$ denotes the angular frequency, $\boldsymbol{u}$ is the particle displacement; $\boldsymbol{\epsilon}$, $\boldsymbol{\tau}$ are the 2nd-order strain and stress tensors; $\boldsymbol{c}$ is the 4th-order elastic tensor. For Voigt notation, $c_{ijkl}$ can be converted to second-order tensor $(C_{rs})_{6\times6}$. The subscripts of $\boldsymbol{C}$ and $\boldsymbol{c}$ satisfy the relations between $(r,s)$ and $(i,j,k,l)$: $1\leftrightarrow11$, $2\leftrightarrow22$, $3\leftrightarrow33$, $4\leftrightarrow23$, $5\leftrightarrow13$ and $6\leftrightarrow12$. Therefore, the constitutive equation \eqref{tau} can be transformed into a matrix form \cite{conry2005notes}. And, the divergence of $\boldsymbol{\tau}$ computed by the left divergence operator can be expressed as
\begin{equation}\label{5}
	\nabla\cdot\boldsymbol{\tau}=\hat{\bf{e}}_k\partial x_k\cdot\tau_{ji}\hat{\bf{e}}_j\otimes\hat{\bf{e}}_i=\sum\limits_{j=1}^{3}\frac{\partial\tau_{ji}}{\partial x_j}\hat{\bf{e}}_i\triangleq\tau_{ji,j}\hat{\bf{e}}_i
	\end{equation}	
where Einstein's convention has been adapted, with the repeated indices implying summation.  The operators ``$\otimes$" and ``$\cdot$`` represent the diadic product the dot product, respectively. Substituting \eqref{5} into \eqref{wave equation} yields
	\begin{equation}\label{4}
	\omega^2\rho_{ij}u_j(x,y,z)+\tau_{ji,j}=0
	\end{equation}
The Latin subscripts $i,j,k,\dots$  represent three-dimensional indices and the Greek subscripts $\alpha,\beta,\dots$ are two-dimensional indices. It is well known that the waveguide problem is actually a 2.5-dimensional problem, where the field is three-dimensional depending on $(x_1,x_2,x_3)=(x,y,z)$ but the materials are two-dimensional depending on $(x_1,x_2)$. When the propagation is along the $+z$-axis and the cross section of the waveguide is uniform in the $z$-direction, the phasor expression for displacement field $\boldsymbol{u}$ and the operator $\boldsymbol{\nabla}$ can be written explicitly as
\begin{subequations}
\begin{align}
\label{2}\boldsymbol{u}&=\hat{\bf{e}}_iu_i(x,y)e^{-\gamma_zz}e^{\mathrm{j}\omega t}\\
\label{3}\boldsymbol{\nabla}&=\hat{\bf{e}}_1\frac{\partial}{\partial x_1}+\hat{\bf{e}}_2\frac{\partial}{\partial x_2}-\hat{\bf{e}}_3\mathrm{j}\mathrm{k}_z\equiv\boldsymbol{\nabla}_t-\hat{\bf{e}}_3\gamma_z
\end{align}
\end{subequations}
for any given waveguide mode, where $\hat{\bf{e}}_i$ and $u_i(x,y)$ denote the $i$-th unit vector and its corresponding component of the displacement field in Cartesian coordinates, respectively, and $\gamma_z=\mathrm{j}\mathrm{k}_z =\alpha_z+\mathrm{j}\beta_z$ is the complex propagation constant  (the real variables $\alpha_z$ and $\beta_z$ are called the attenuation constant and phase constant respectively). $\mathrm{k}_z$ is the $z$-component of the wave vector. In the following formulations, the time convention $e^{\mathrm{j}\omega t}$ is omitted. Therefore, substituting (4) into \eqref{epsilson}, the strain tensor can be written as
\begin{equation}\label{6}
\boldsymbol{\epsilon}=e^{-\gamma_zz}\left[\begin{array}{ccc}
u_{1,1}& \dfrac{u_{1,2}+u_{2,1}}{2}&\dfrac{u_{3,1}-\gamma_zu_1}{2}\\
\dfrac{u_{1,2}+u_{2,1}}{2}&u_{2,2}&\dfrac{u_{3,2}-\gamma_zu_2}{2}\\
\dfrac{u_{3,1}-\gamma_zu_1}{2}&\dfrac{u_{3,2}-\gamma_zu_2}{2}&-\gamma_zu_3
\end{array}\right]
	\end{equation}
On the other hand, the stress tensor can be expressed as $\boldsymbol{\tau}=e^{-\gamma_zz}\tau_{ij}\hat{\bf{e}}_i\otimes\hat{\bf{e}}_j$, and the components $\tau_{ij}$ are indicated as
\begin{equation}\label{tauij}
\tau_{ij} = C_{r(i,j),r(k,\alpha)}u_{k,\alpha}
-\gamma_zC_{r(i,j),r(k,3)}u_k
\end{equation}
where the subscripts of $C$ come from the elements of a symmetrical constant matrix $r$ defined by
$$ r=\left[\begin{array}{ccc}1&6&5\\6&2&4\\5&4&3\end{array}\right]$$
Obviously, $\boldsymbol{\tau}$ is also symmetrical. Inserting (4)-\eqref{tauij} into \eqref{4}, we can obtain the governing equation of the elastic waveguide as follows
\begin{itemize}
\item  Tensor formulation:
\begin{align}\label{main equ}
\gamma_z^2\ell_{3ij3}\hat{\bf{e}}_i\otimes\hat{\bf{e}}_j\cdot\boldsymbol{u}-\gamma_z[\boldsymbol{\nabla_t}\cdot &(\ell_{\alpha ij3}\hat{\bf{e}}_\alpha\otimes\hat{\bf{e}}_i\otimes\hat{\bf{e}}_j \cdot\boldsymbol{u})+\ell_{3ij\alpha}\hat{\bf{e}}_i\otimes\hat{\bf{e}}_j\otimes\hat{\bf{e}}_\alpha:\boldsymbol{\nabla_t}\boldsymbol{u}]+\notag\\
&\boldsymbol{\nabla_t}\cdot(\ell_{\alpha ij\beta}\hat{\bf{e}}_\alpha\otimes\hat{\bf{e}}_i\otimes\hat{\bf{e}}_j\otimes\hat{\bf{e}}_\beta:\boldsymbol{\nabla_t}\boldsymbol{u})+\omega^2\boldsymbol{\rho}\cdot\boldsymbol{u}=0
\end{align} 
\footnotetext[1]{$\hat{\bf{e}}_i\otimes\hat{\bf{e}}_j\cdot\hat{\bf{e}}_k=\hat{\bf{e}}_i\delta_{jk}$, $\hat{\bf{e}}_i\cdot\hat{\bf{e}}_j\otimes\hat{\bf{e}}_k=\delta_{ij}\hat{\bf{e}}_k$ and $\hat{\bf{e}}_i\otimes\hat{\bf{e}}_\alpha:\hat{\bf{e}}_\beta\otimes\hat{\bf{e}}_j=\delta_{ij}\delta_{\alpha\beta}$}
\footnotetext[2]{For a given $k$, for example $k=3$, $\ell_{3ijm}\hat{\bf{e}}_3\otimes\hat{\bf{e}}_i\otimes\hat{\bf{e}}_j\otimes\hat{\bf{e}}_m$ is a 3rd-order tensor. Thus, we write it as $\ell_{3ijm}\hat{\bf{e}}_i\otimes\hat{\bf{e}}_j\otimes\hat{\bf{e}}_m$ for short.}
\item Component formulation: 
\begin{equation}\label{main equ2}
\gamma_z^2\ell_{3ij3}u_j-\gamma_z[\partial x_\alpha(\ell_{\alpha ij3}u_j)+\ell_{3ij\alpha}u_{j,\alpha}]+\partial  x_\alpha(\ell_{\alpha ij \beta}u_{j,\beta})+\omega^2\rho_{ij}u_j=0
\end{equation}
\item The components of the coefficient tensor $\boldsymbol{\ell}=\ell_{kijm}\hat{\bf{e}}_k\otimes\hat{\bf{e}}_i\otimes\hat{\bf{e}}_j\otimes\hat{\bf{e}}_m$ are denoted by
\begin{equation}\ell_{kijm}=C_{r(i,k),r(j,m)}\end{equation}\footnotetext[3]{$\ell_{kijm}=\ell_{ikjm}=\ell_{kimj}=\ell_{jmki}$}
\end{itemize}
It is easy to see that \eqref{main equ2} is a quadratic eigenvalue problem, where $\gamma_z=\mathrm{j}k_z=\alpha_z+\mathrm{j}\beta_z$ is the eigenvalue and $\boldsymbol{u}$ denotes the corresponding eigenvector. The goal of this work is to develop the SEM for solving the eigenpairs $(\gamma_z,\boldsymbol{u})$.

\subsection{Weak Formulation}
Based on the framework of FEM, multiplying equation \eqref{main equ} by the test function $\varphi$ and integrating and using the integration by parts for the second and fourth integrations, we arrive at the weak form equation for the solid region
 \begin{equation}\label{main inte}
 \gamma_z^2a_{ij}(u_j,\varphi)-\gamma_z [b_{ij}(u_j,\varphi)+I_1]+q_{ij}(u_j,\varphi)+I_2=0
 \end{equation}
The above bilinear functions can be  written in detail as
\begin{subequations}
\begin{align}
&a_{ij}(u_j,\varphi)=\displaystyle{\int_\Gamma \varphi^\dagger (\ell_{3ij3}u_j)dxdy}\\
&b_{ij}(u_j,\varphi)=\displaystyle{\int_\Gamma \varphi^\dagger (\ell_{3ij\alpha}u_{j,\alpha})-(\partial x_\alpha\varphi)^\dagger(\ell_{\alpha ij3}u_j)dxdy}\\
&q_{ij}(u_j,\varphi)=\displaystyle{\int_\Gamma\omega^2\varphi^\dagger(\rho_{ij}u_j)-(\partial x_\alpha \varphi)^\dagger(\ell_{\alpha ij \beta}u_{j,\beta})dxdy}\\
&I_1=\displaystyle{\int_{\partial\Gamma}\varphi^\dagger(n_{\alpha}\ell_{\alpha ij3} u_j)dx}\label{bd3}\\
&I_2=\displaystyle{\int_{\partial\Gamma}\varphi^\dagger(n_{\alpha}\ell_{\alpha ij \beta}u_{j,\beta})dx}\label{bd2}
\end{align}
\end{subequations}
 where $\Gamma$ is the cross section of waveguide; $\partial\Gamma$ denotes the boundary of $\Gamma$; $\hat{\mathbf{n}}=n_\alpha \hat{\bf{e}}_\alpha$ represents the unit outward normal vector  at the point on the edge $\partial\Gamma$; the superscript ``$\dagger$" means  the complex conjugate. Because of the existence of boundary integral items $I_{1}$ and $I_{2}$, we need one or the combination of the following boundary conditions.
 \subsection{Boundary Conditions}
 In order to solve the propagation constant $\gamma_z$ within a given waveguide, for the external boundary integral $-\gamma_zI_1+I_2$, we need the suitable boundary conditions, such as the hard BC, the soft BC, the BPBC and the ABC.
 \begin{enumerate}[(1)]
 \item Hard BC reads $\boldsymbol{u}=0$. Therefore $-\gamma_zI_1+I_2=0$ and the region of the integration $\Gamma$ is replaced by $\Gamma\backslash\partial\Gamma$.
 \item Soft BC reads ${\hat{\textbf n}}\cdot\boldsymbol{\tau}=0$.  According to \eqref{tauij}, we have
\begin{align}
n_\alpha\tau_{\alpha i}=n_\alpha [-\gamma_z(\ell_{\alpha ij3}u_j)+\ell_{\alpha ij\beta}u_{j,\beta}]\label{tau2j}
\end{align}
Substituting \eqref{tau2j} into the external boundary integral, we have $-\gamma_zI_1+I_2=\int_{\partial \Gamma}{\hat{\textbf n}}\cdot\boldsymbol{\tau} dx$  in view of ${\hat{\textbf n}}\cdot\boldsymbol{\tau}=n_j\tau_{ji}$ and 
$n_3=0$ in the waveguide problem. Therefore, the boundary integrals vanish and the region of the integration is unchanged.
\item BPBC.
 \begin{enumerate}[(a)]
 \item By the Bloch theorem \cite{niu2017spectral}, we first obtain
 \begin{equation}
 \label{bloch the}
 u_j(\mathbf{k}_t,\boldsymbol{r}+\boldsymbol{a})= u_j(\mathbf{k}_t,\boldsymbol{r})e^{-\mathrm{j}\mathbf{k}_t\cdot\boldsymbol{a}}
 \end{equation}
where $ \bold{k} = \bold{k}_t + \hat{z}\mathrm{k}_z $ is the Bloch wave vector, $\boldsymbol{r}$ and $ \boldsymbol{a}$ are the position vectors on the boundary $\partial\Gamma$ and the lattice translation vector, respectively. 
Define the Bloch periodic subspace
\begin{equation}
 H_p^B(\Gamma)=\{v\in H^1(\Gamma):v(\boldsymbol{r}+\boldsymbol{a})=v(\boldsymbol{r})e^{-\mathrm{j}\mathbf{k}_t\cdot\boldsymbol{a}}~ \text{on}~\partial \Gamma \}
 \end{equation}
 where $H^1(\Gamma)=\{v\in L_2(\Gamma):\boldsymbol{\nabla_t} v\in L_2(\Gamma)^2\}$.
  For any $\varphi$, $u_{j}$ belonging to $H_{p}^{B}$, from \eqref{bd3} and \eqref{bd2}, note that the normal vectors defined on a pair of periodic boundary have opposite directions, we can obtain the external boundary integral  $I_1=I_2=0$. Consequently, the external boundary integrations vanish for all the opposite boundaries by using the BPBC on the external boundary.
  \item Meanwhile, the BPBC waveguide problem can be transformed into the equivalent periodic boundary conditions (PBC) waveguide problem. When $u_j$ are written as the plane wave form
$u_j(\mathbf{k}_t,\boldsymbol{r})=\widetilde{u}_j(\mathbf{k}_t,\boldsymbol{r})e^{-\mathrm{j}\mathbf{k}_t\cdot\boldsymbol{r}}$,
 we can obtain the periodic boundary conditions $\widetilde{u}_j(\mathbf{k}_t,\boldsymbol{r}+\boldsymbol{a})=\widetilde{u}_j(\mathbf{k}_t,\boldsymbol{r})$ arising from \eqref{bloch the}. Therefore, the corresponding periodic subspace can be defined by
\begin{equation}
H_p(\Gamma)=\{v\in H^1(\Gamma):v(\boldsymbol{r}+\boldsymbol{a})=v(\boldsymbol{r})~\text{on}~ \partial \Gamma \}\end{equation}
For any $\varphi$, $u_{j}$ belonging to $H_{p}$, it is easy to check that the external boundary integrals are still zero. By replacing the operator $\boldsymbol{\nabla_t}$ with $\boldsymbol{\nabla_t}-\mathrm{j}\mathbf{k}_t$ in \eqref{main inte}, we arrive at a new scheme
\begin{equation}\label{main inte2}
\gamma_z^2\widetilde{a}_{ij}(\widetilde{u}_j,\varphi)-\gamma_z\widetilde{b}_{ij}(\widetilde{u}_j,\varphi)+\widetilde{q}_{ij}(\widetilde{u}_j,\varphi)=0
\end{equation}
 where the bilinear functions can be written as
\begin{align}
&\widetilde{a}_{ij}(\widetilde{u}_j,\varphi)=\displaystyle{\int_\Gamma \varphi ^\dagger(\ell_{3ij3}\widetilde{u}_j)dxdy}\notag\\
&\widetilde{b}_{ij}(\widetilde{u}_j,\varphi)=\displaystyle{\int_\Gamma \varphi^\dagger [\ell_{3ij\alpha}(\widetilde{u}_{j,\alpha}-\text{j}\text{k}_\alpha\widetilde{u}_j)]-(\partial x_\alpha\varphi-\text{j}\text{k}_\alpha)^\dagger(\ell_{\alpha ij3}\widetilde{u}_j)dxdy}\\
&\widetilde{q}_{ij}(\widetilde{u}_j,\varphi)=\displaystyle{\int_\Gamma\omega^2\varphi^\dagger\rho_{ij}\widetilde{u}_j-(\partial x_\alpha \varphi-\text{j}\text{k}_\alpha)^\dagger[\ell_{\alpha ij \beta}(\widetilde{u}_{j,\beta}-\text{j}\text{k}_\beta\widetilde{u}_j)]}dxdy\notag
\end{align}
\end{enumerate}
\item ABC.\\
In our work, the ABC is used to truncate the infinite external boundary when the waveguide is unbounded in the transverse directions. From \cite{Komatitsch1999The}, the ABC for the scalar mass density is expressed as
\begin{equation}\label{abc}\boldsymbol{t}=C_L\rho(\boldsymbol{v}\cdot\hat{\mathbf{n}})\hat{\mathbf{n}}+C_T\rho(\boldsymbol{v}-(\boldsymbol{v}\cdot\hat{\mathbf{n}})\hat{\mathbf{n}})\end{equation} where $\boldsymbol{t}$ is the boundary traction, $\boldsymbol{v}$ is the velocity field on the surface, and $C_L$, $C_T$ represent the material bulk speed of longitudinal and transverse waves in the background outside the computational region $\Gamma$, respectively. However, in this work we treat anisotropic density, so the boundary function is rewritten as
\begin{equation}
\hat{\mathbf{n}}\cdot\boldsymbol{\tau}=\mathrm{j}\omega\boldsymbol{\rho}\cdot[C_T\boldsymbol{u}+(C_L-C_T)(n_ju_j)\hat{\mathbf{n}}]
\end{equation}
Accordingly, the scalar expressions with the Einstein notation are expressed as
\begin{equation}
\mathrm{n}_j\tau_{ji}=\mathrm{j}\omega\rho_{ij}[C_Tu_j+(C_L-C_T)(\mathrm{n}_1u_1+\mathrm{n}_2u_2)\mathrm{n}_j]
\end{equation}
Moreover, noting that $\mathrm{n}_3=0$, the external boundary integrations in \eqref{main inte} are replaced by
\begin{equation}
-\gamma_z I_1+I_2= d_{ij}(u_j,\varphi)
\end{equation}
The above bilinear function can be written in detail as 
\begin{equation}
d_{ij}(u_j,\varphi)=\mathrm{j}\omega \int_{\partial \Gamma_{\text{ext}}}\varphi ^\dagger[\rho_{ij}C_T+(C_L-C_T)\mathrm{n}_j\rho_{i\alpha}\mathrm{n}_\alpha]u_jdx
\end{equation}
The weak formulations of the elastic waveguide for the pure solid model with the BPBC and the ABC are expressed compactly as
 \begin{subequations}
 \begin{align}
 &\gamma_z^2a_{ij}(u_j,\varphi)-\gamma_z b_{ij}(u_j,\varphi)+q_{ij}(u_j,\varphi)+I_3=0\\
 &I_3=\left\{\begin{array}{ll}
 0 &\text{for Hard BC, Soft BC and BPBC}\\
 d_{ij}(u_j,\varphi) ~&\text{for ABC}
 \end{array}
 \right.\label{solid ext}
 \end{align}
 \end{subequations}
 \end{enumerate}
The above is for the case where all materials are solid in the waveguide.  If part of the waveguide is made of fluid, we need to consider the special coupling between fluid and solid in the eigenvalue problem.

\subsection{The Fluid-Solid Coupling System}
\begin{figure}[!htbp]
\centering
\begin{overpic}[scale=0.22]%
               {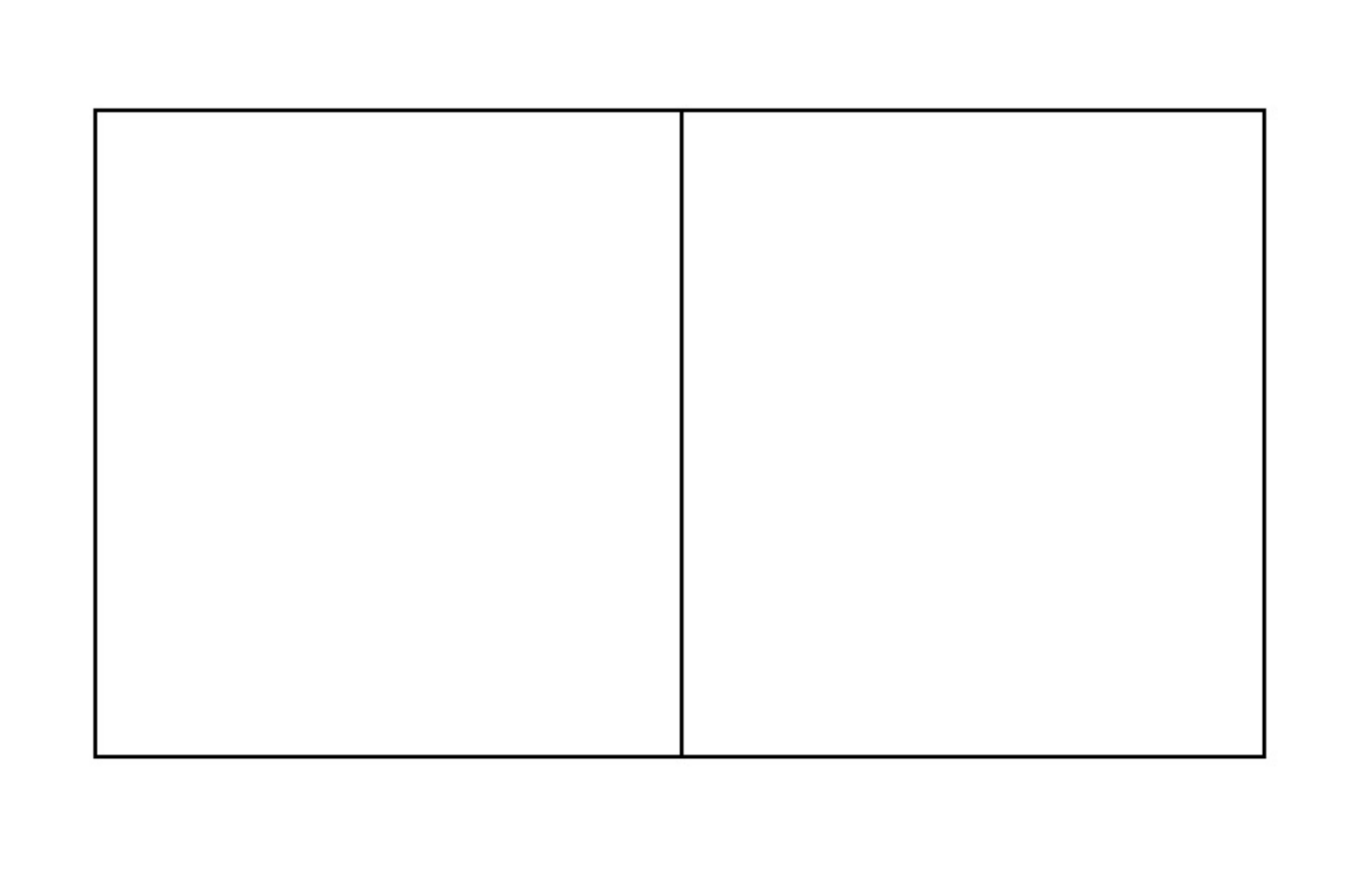}
              \put(20,30){solid}
              \put(65,30){fluid}
               \put(44,30){$\partial\Gamma_{fs}$}
\end{overpic}
\caption{The solid region $s$ (left) and the fluid region $f$ (right) with an interface $\partial\Gamma_{fs}$.}
\label{solidfluid}
\end{figure}

When the waveguide is filled with an inhomogeneous medium including parts of fluid and solid, we need to consider the fluid-solid coupling. The symbol ``$f$'' is introduced to denote the local fluid region $f$, which shares the common interface $\partial\Gamma_{fs}$ with the local solid region $s$ shown in Fig \ref{solidfluid}. For the fluid-solid coupling system, we not only derive the weak form in the fluid region, but also give the continuity condition of the fluid-solid at the interface defined by $\partial \Gamma_{fs}$.
 First, the governing equation for the potential $\chi$, defined as $\boldsymbol{v}=\rho_f^{-1}\boldsymbol{\nabla}\chi$, in the fluid region is introduced from \cite{komatitsch2000wave}
\begin{equation}\label{fluidgove}\boldsymbol{\nabla}\cdot(\rho_f^{-1}\boldsymbol{\nabla}\chi)+\kappa^{-1}\omega^2\chi=0\end{equation}
where $\rho_f$ is the density of fluid/acoustic materials and $\kappa$ is the bulk modulus. The phasor expression for potential is shown as $\chi=\chi(x,y)e^{-\gamma_zz}$.
Multiplying \eqref{fluidgove} by the test function $\psi$ and integrating, after using the integration by parts, we obtain the weak form for the fluid region 
\begin{equation}\label{fluid inte}
\gamma_z^2 e(\chi,\psi)-f(\chi,\psi)+I_4=0
\end{equation}
The above bilinear functions can be expressed as follows
\begin{align}
&e(\chi,\psi)=\displaystyle{\int_\Gamma\psi^\dagger\rho_f^{-1}\chi dxdy}\tag{\ref{fluid inte}{a}}\\
&f(\chi,\psi)=\displaystyle{\int_\Gamma(\partial_\alpha\psi)^\dagger(\rho_f^{-1}\partial_\alpha\chi)-\psi^\dagger\kappa^{-1}\omega^2\chi dxdy}\tag{\ref{fluid inte}{b}}\\
&I_4=\displaystyle{\int_{\partial\Gamma}\psi^\dagger (n_\alpha \rho_f^{-1}\partial_\alpha\chi) dx}\label{I4}\tag{\ref{fluid inte}{c}}
\end{align}
Second, when the boundary integral $-\gamma_zI_{1}+I_{2}$ in \eqref{main inte} is restricted to the interface $\partial\Gamma _{fs}$ between the solid region and the fluid region, because of the continuity condition of the traction $\hat{\mathbf{n}}\cdot \boldsymbol{\tau} =\mathrm{j}\omega\chi\hat{\mathbf{n}}$ \cite{komatitsch2000wave} and the fact $\mathrm{n}_3=0$, it follows that
\begin{eqnarray}
-\gamma_zI_1^{fs}+I_2^{fs}=\displaystyle{\int_{\partial \Gamma_{fs}}\varphi^\dagger \mathrm{j}\omega\chi(x,y)\mathrm{n}_i dx}\triangleq  I_5^{fs}\end{eqnarray}
Finally, for the fluid region, similarly, the external boundary integration $I_{4}$ satisfies
\begin{equation}
\label{fliud ex}
I_4=\left\{\begin{array}{ll}
0 &\text{for Hard BC, Soft BC, and BPBC}\\
 \mathrm{j}\omega\displaystyle{\int_{\partial\Gamma_{\text{ext}}}\psi^\dagger(\rho_fC_L)^{-1}\chi dx}&\text{for ABC}
\end{array}\right.
\end{equation}
On the other hand, by replacing the normal component of the velocity $\hat{\mathbf{n}}\cdot\boldsymbol{v}_{\text{fluid}}=\hat{\mathbf{n}}\cdot(\rho_f^{-1}\boldsymbol{\nabla}\chi)$ in the fluid region with the normal component of the velocity $\hat{\mathbf{n}}\cdot(\mathrm{j}\omega\boldsymbol{u})$ in the solid region \cite{komatitsch2000wave},  $I_4^{fs}$ can be derived as
\begin{equation} I_4^{fs}=\mathrm{j}\omega\int_{\partial \Gamma_{fs}}\psi^\dagger\mathrm{n}_ju_j dx\end{equation}
Compactly, the weak formulations of the fluid-solid coupling system are shown as
\begin{equation}
\label{fsequ}\left\{
\begin{array}{l}\gamma_z^2a_{ij}(u_j,\varphi)-\gamma_z[b_{ij}(u_j,\varphi)+I_1]+q_{ij}(u_j,\varphi)+I_2+I_5^{fs}=0\\
\gamma_z^2 e(\chi,\psi)-f(\chi,\psi)+I_4+I_4^{fs}=0
\end{array}
\right.
\end{equation}
Note that, when the cladding medium outside a core of the waveguide is filled with solid, $-\gamma_zI_1+I_2$ is replaced by $I_3$ shown in \eqref{solid ext} and $I_4=0$. Conversely, when the cladding is a fluid region, $-\gamma_z I_1+I_2=0$ and $I_4$ is shown in \eqref{fliud ex}.\\
The above completes the formulation of elastic waveguide and its weak formulations. In the next section, we will introduce the discretization scheme to calculate the propagation constants $\gamma_z$ of the waveguide and their corresponding modes (eigenvectors).

\section{BASIS FUNCTIONS AND DISCRETIZATION}
\subsection{Basis Functions}
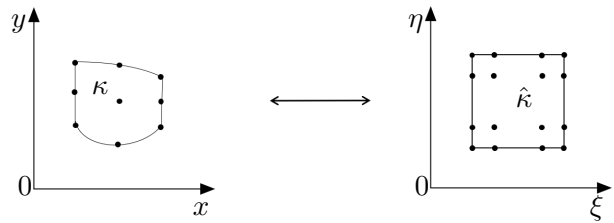
\begin{figure}[!htbp]
\centering
\begin{overpic}[scale=0.3]
               {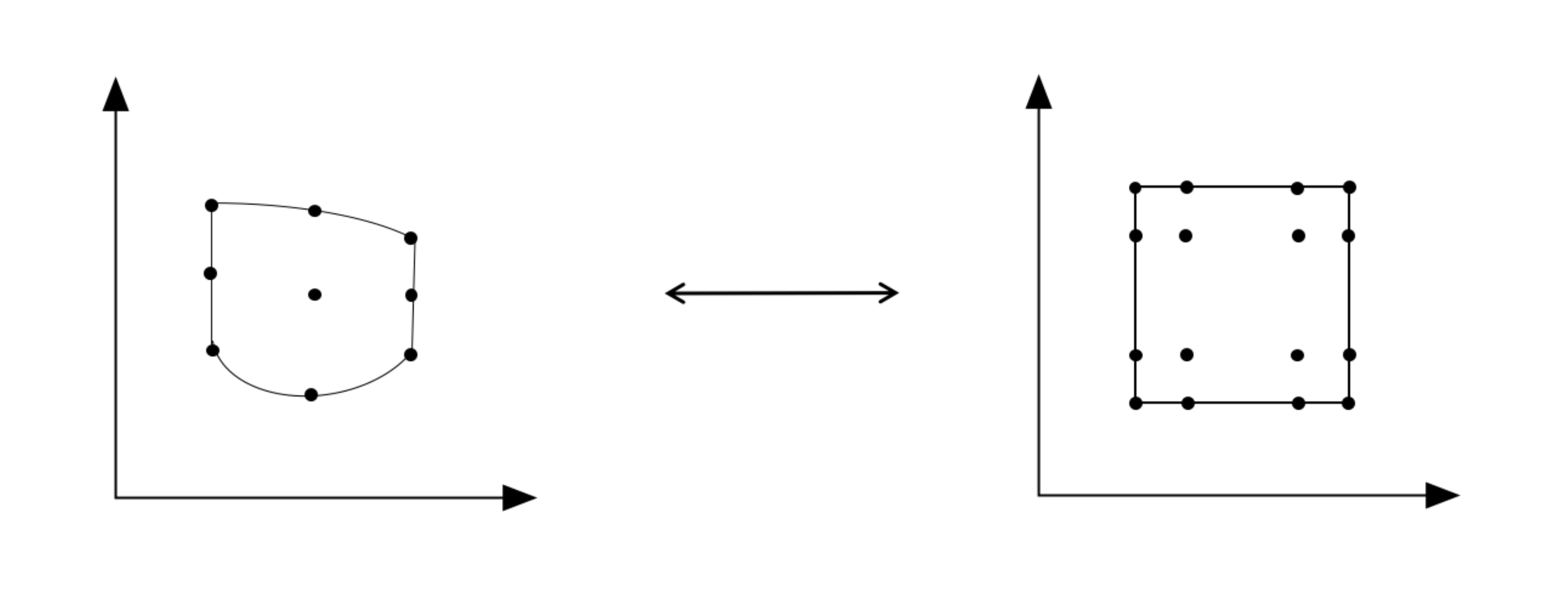}
               \put(31,2){$x$}
               \put(4,30){$y$}
               \put(5,5){0}
            \put(64,5){0}
\put(16,20){$\kappa$}
\put(79,18){$\hat{\kappa}$}
               \put(90,2){$\xi$}
               \put(63,30){$\eta$}
\end{overpic}
\caption{The 3rd-order curvilinear mapping between $\kappa$ and $\hat{\kappa}$. (Left) A second-order geometrical curved element $\kappa$ in the physical domain. (Right) The corresponding reference element $\hat{\kappa}$ for the 3-rd SEM ($N=3$), where the 16 points are GLL points.}
\label{map}
\end{figure}

In order to approximate the unknown field component $u_j$, we apply the GLL polynomials as the basis functions. The $N$th-order 1D GLL polynomials are defined as
 \begin{equation}
 \phi_r^{(N)}=\dfrac{-1}{N(N+1)L_N(\xi_r)}\dfrac{(1-\xi^2){L}_N^{'}(\xi)}{(\xi-\xi_r)}, r=1,2,\cdots,N+1
 \end{equation}
 where the interpolating points $\xi_r\in[-1,1]$, and they are chosen as the GLL points which are the roots of equation $(1-\xi_r^2)L_N'(\xi_r)=0$, and $L_N'(\xi)$ is the derivative of the $N$th-order Legendre polynomial. Note that, the Legendre polynomials are orthogonal polynomials that allow to reduce the interpolation errors compared to the standard Lagrange polynomials used in the FEM.
 $u_j$ can be approximated by using the tensor-product $\varphi_p^{(N)}=\phi_r^{(N)}(\xi)\phi_s^{(N)}(\eta)$ of two 1D nodal basis functions, where the subscript $p$ is the compound index of $(r,s)$. Let the physical domain be subdivided into a number of non-overlapping quadrilateral elements, so that each element can be mapped into the reference element $[-1,1]\times[-1,1]$ by the mapping $x(\xi,\eta),y(\xi,\eta)$ \cite{LiuMixed,luo2009spectral}. For example, the irregular element $\kappa$ with curved edges can be mapped to the reference element $\hat{\kappa}$ by using the curvilinear mapping shown in Fig \ref{map}.
 While the corresponding invertible mappings are applied to the basis function $\varphi(x,y)=\widehat{\varphi}(\xi,\eta)$ and $\boldsymbol{\nabla_t}\varphi(x,y)=\mathbf{J}^{-1}\widehat{\boldsymbol{\nabla_t}}\widehat{\varphi}(\xi,\eta)\triangleq \mathcal{J}_\alpha\hat{\varphi}\hat{\bf{e}}_\alpha$, where $\mathbf{J}=\left[\begin{array}{cc} \frac{\partial x}{\partial \xi} &\frac{\partial y}{\partial \xi}\\
\frac{\partial x}{\partial \eta} &\frac{\partial y}{\partial \eta}\end{array}\right]$ is the Jacobian matrix, as derived in \cite{ShiSpectral,1542240,LeeA}.

\subsection{Discrete Forms}
In general, three unknown components  $u_j$ of the displacement field can be approximated by
 \begin{equation}\label{disuj}u_{j}=\sum\limits_{q=1}^{\mathrm{n}_{sj}}u_{j,q}\varphi_q^{(N)}(x,y)\end{equation}
where $\mathrm{n}_{sj}$ represents the number of nodal degrees of freedom (DOF) of the component $u_j$ for the solid region. Thus, the total number of DOF in the solid region is $N_s=\sum\limits_{j=1}^{3} N_{sj}$. Inserting \eqref{disuj} into (18), we arrive at the quadratic eigenvalue problems
\begin{eqnarray}
\label{linear2}&[\gamma_z^2\bar{\bar{A}}^{s}-\gamma_z\bar{\bar{B}}^{s}+(\bar{\bar{K}}^{s}+\bar{\bar{M}}^{s})+\bar{\bar{T}}^{s}]{\textbf u}=0
\end{eqnarray}
where $\boldsymbol{u}\triangleq[\boldsymbol{u}_1,\boldsymbol{ u}_2,\boldsymbol{ u}_3]^T$, ${\textbf u}_j\triangleq(u_{j,1},\cdots ,u_{j,\mathrm{n}_{sj}})$, the subscript ``$s$'' means the solid region and $\bar{\bar{T}}$ is the boundary integral matrix, which is equal to zero when using the Hard BC, the Soft BC and the BPBC, and nonzero for the ABC.
 After the invertible mapping, the elemental matrices consist of the following parts
\begin{equation}
\begin{array}{l}
\label{solid element}
(\bar{\bar{A}}_{ik}^{(\hat{\kappa})})_{pq}=\displaystyle{\int_{-1}^1\int_{-1}^1|\boldsymbol{J}| \hat{\varphi}^\dagger_p\ell_{3ik3}\hat{\varphi}_q~d\xi d\eta}\\
(\bar{\bar{B}}_{ik}^{(\hat{\kappa})})_{pq}=\displaystyle{\int_{-1}^1\int_{-1}^1|\boldsymbol{J}|\hat{\varphi}_p^\dagger (\ell_{3ik\alpha}\mathcal{J}_\alpha\hat{\varphi}_q)-(\mathcal{J}_\alpha\hat{\varphi}_p)^\dagger(\ell_{\alpha ik3}\hat{\varphi}_q)d\xi d\eta}\\
(\bar{\bar{K}}_{ik}^{(\hat{\kappa})})_{pq}=-\displaystyle{\int_{-1}^1\int_{-1}^1|\boldsymbol{J}|(\mathcal{J}_\alpha \hat{\varphi}_p)^\dagger(\ell_{\alpha ik \beta}\mathcal{J}_\beta\hat{\varphi}_q)d\xi d\eta}\\
(\bar{\bar{M}}_{ik}^{(\hat{\kappa})})_{pq}=\omega^2\displaystyle{\int_{-1}^1\int_{-1}^1|\boldsymbol{J}|\rho_{ik}\hat{\varphi}_p^\dagger\hat{\varphi}_qd\xi d\eta}\\
(\bar{\bar{T}}_{ik}^{(\hat{\kappa})})_{pq}=\mathrm{j}\omega\displaystyle{\int_{-1}^1|\boldsymbol{J}_b|\hat{\varphi}_p^\dagger[\rho_{ik}C_T+(C_L-C_T)\mathrm{n}_k\rho_{i\alpha}\mathrm{n}_\alpha]\hat{\varphi}_qd\xi}

\end{array}
\end{equation}
where $p,q=1,2\cdots,N+1$, the superscript ``$(\hat{\kappa})$'' means the reference element and $\boldsymbol{J}_b$ arises from the mapping from any edges to reference domain $[-1,1]$. Meanwhile, $\bar{\bar{T}}^{s}$ arising from the ABC is expressed in \eqref{solid element}.
Similarly, $u_j,\chi$ are written as
\begin{equation}\label{dis2}
u_j^{(s)}=\sum\limits_{q=1}^{\mathrm{n}_{sj}}u_{j,q}^{(s)}\varphi_q^{(s)}
,\quad 
w^{(f)}=\sum\limits_{q=1}^{\mathrm{n}_{f}}w_q^{(f)}\varphi_q^{(f)}
\end{equation}
where $\mathrm{n}_{f}$ denotes the number of the total nodal DOF for the fluid region. Substituting (\ref{dis2}) into (\ref{fsequ}), we arrive at the fluid-solid coupling eigenvalue problem based on the BPBC
\begin{align}
\{\gamma_z^2\left[\begin{array}{cc}
                \bar{\bar{A}}^s&0\\
                 0&A_0^f\end{array}\right]-\gamma_z\left[\begin{array}{cc}
                                                       \bar{\bar{B}}^s&0\\
                                                        0&0
                                                         \end{array}\right]+\left[\begin{array}{cc}
                                                         \bar{\bar{K}}^s+\bar{\bar{M}}^s&\bar{\bar{R}}^{(s,f)}\\
                                                         \bar{\bar{Q}}^{(f,s)}&K_0^{f}+M_0^{f}
                                                                                  \end{array}\right]\}\left(\begin{array}{c}\boldsymbol{ u}\\\boldsymbol{ w} \end{array}\right)=0\label{eigen_pro_2}
\end{align}
where $\bar{\bar{R}}^{(s,f)}=[\bar{\bar{R}}^{(s,f)}_1, \bar{\bar{R}}^{(s,f)}_2 ,\bar{\bar{R}}^{(s,f)}_3]^T$, $\bar{\bar{Q}}^{(f,s)}=[\bar{\bar{Q}}^{(f,s)}_1, \bar{\bar{Q}}^{(f,s)}_2, \bar{\bar{Q}}^{(f,s)}_3]$, $\boldsymbol{w}\triangleq(w_1,\cdots,w_{\mathrm{n}_{f}})$ . The elemental matrices are given as following
\begin{equation}
\begin{array}{l}
 (A_0^{(f)})_{pq}=\displaystyle{\int_{-1}^1\int_{-1}^1|\boldsymbol{J}| \hat{\varphi}_p^{(f)\dagger}\rho_f^{-1}\hat{\varphi}_q^{(f)} d\xi d\eta}\\
 (K_0^{(f)})_{pq}=-\displaystyle{\int_{-1}^1\int_{-1}^1|\boldsymbol{J}| (\mathcal{J}_\alpha\hat{\varphi}_p^{(f)})^\dagger\rho_f^{-1}(\mathcal{J}_\alpha\hat{\varphi}_q^{(f)}) d\xi d\eta}\\
 (M_0^{(f)})_{pq}=\displaystyle{\int_{-1}^1\int_{-1}^1|\boldsymbol{J}|\hat{\varphi}_p^{(f)\dagger}\kappa^{-1}\omega^2\hat{\varphi}_q^{(f)} d\xi d\eta}\\
 (\bar{\bar{Q}}^{(f,s)}_i)_{pq}=\displaystyle{\mathrm{j}\omega\int_{-1}^1|\boldsymbol{J}_b|\hat{\varphi}_p^{(f)\dagger}\mathrm{n}_i\hat{\varphi}_q^{(s)} d\xi}\\
(\bar{\bar{R}}^{(s,f)}_i)_{pq}=\mathrm{j}\omega\displaystyle{\int_{-1}^1|\boldsymbol{J}_b|\hat{\varphi}_p^{(s)\dagger}\hat{\varphi}_q^{(f)}\mathrm{n}_id\xi}
 \end{array}
 \end{equation}
For the remaining elemental matrices, they can be obtained by replacing the superscript $(\hat{\kappa})$ of \eqref{solid element} with $(s)$. After these matrices are assembled, the quadratic eigenvalue problems \eqref{eigen_pro_2} is converted to a first order generalized eigenvalue problem in \eqref{1st_eigv} for $\gamma_z$ referring to \cite{coise2000backward},
\begin{equation}\label{1st_eigv}
\left[\begin{array}{cccc}
\bar{\bar{B}}^s& \mathbf{0}&-\bar{\bar{K}}^s-\bar{\bar{M}}^s&-\bar{\bar{R}}^{(s,f)}\\
 \mathbf{0}& \mathbf{0}&-\bar{\bar{Q}}^{(f,s)}&-K_0^{f}-M_0^{f}\\
  \mathbf{I}& \mathbf{0}& \mathbf{0}& \mathbf{0}\\
  \mathbf{0}& \mathbf{I}& \mathbf{0}& \mathbf{0}
\end{array}\right]
\left[\begin{array}{c}
\gamma_z\boldsymbol{u}\\
\gamma_z\boldsymbol{w}\\
\boldsymbol{u}\\
\boldsymbol{w}
\end{array}\right]=\gamma_z\left[\begin{array}{cccc}
 \bar{\bar{A}}^s&\mathbf{0}&\mathbf{0}&\mathbf{0}\\
                \mathbf{0}&A_0^f&\mathbf{0}&\mathbf{0}\\
                \mathbf{0}&\mathbf{0}&\mathbf{I}&\mathbf{0}\\
                \mathbf{0}&\mathbf{0}&\mathbf{0}&\mathbf{I}
\end{array}\right]\left[\begin{array}{c}
\gamma_z\boldsymbol{u}\\
\gamma_z\boldsymbol{w}\\
\boldsymbol{u}\\
\boldsymbol{w}
\end{array}\right]
\end{equation}
 where $\mathbf{I}$ and $\mathbf{0}$ denote the identity matrix and zero matrix, and then it can be solved by using the eigenvalue solver ``eigs" in MATLAB based on ARPACK library routines.
 \section{Numerical Results}
In this section, several examples are presented to verify the high accuracy and efficiency of the SEM for simulating the elastic waveguide problems. The memory, the number of degrees of freedom (DOF) and the accuracy  for our method are compared with the commercial FEM solver COMSOL. Finally, we conduct a numerical experiment on an elastic matematerial (EMM) core which cannot be solved by COMSOL, because of the presence of anisotropic density. Before the experiments, there are some preparations. First, in the simulation of the BPBC waveguide problem, the wave vector is defined by $${\textbf k}=k(\hat{x}\sin{\theta}\cos{\phi}+\hat{y}\sin{\theta}\sin{\phi}+\hat{z}\cos{\theta})$$
where  $k=\omega/v_i$, $v_i$ is the velocity of the P wave or S wave in the background medium and  $(\theta,\phi)$ are the elevation and azimuthal angles of the propagation direction. Second, for convenience, we introduce the notations in our tables and figures: 1) $v_{p}$ is the velocity of the P wave (longitudinal wave). 2) $v_{s} $ is the velocity of the S wave (transversal waves).
3) $\rho$ is the mass density. 
4) $\lambda$ and $\mu$ are Lam\`{e} constants.
5) k$_{i,z}^N$ is the i-th eigenmode wavenumber $\mathrm{k}_z$ obtained in the $z$ direction by the $N$-th order SEM.
6) The reference value $\bar{\mathrm{k}}_{i,z}^{10}$ is the solution of the 10th-order SEM with an extremely fine mesh. The relative error is calculated by $|\mathrm{k}_{i,z}^{N}-\bar{\mathrm{k}}_{i,z}^{10}|/|\bar{\mathrm{k}}_{i,z}^{10}|$. 7) The computational time and memory are displayed with the ``tic", ``toc" function and ``memory" function in Matlab, respectively. Last but not least, for the quadratic eigenvalue problems, the solver will provide two opposite eignvalues ($\gamma_z$ and $-\gamma_z$). To determine the correct sign of the propagation constant, we introduce two quantities: the time averaged Poynting vector \cite{seshadri1983energy} $\boldsymbol{p}=\text{Re}(-\mathrm{j}\omega\boldsymbol{ u}\cdot\boldsymbol{\tau}^\dagger)/2$ and the corresponding power $P_z$=$\int_{\Gamma}\hat{\textbf z}\cdot \boldsymbol{p} dxdy$ in the cross section. The positive $P_z$ is the criterion for {\color{re2}{choosing}} the correct sign of $\gamma_z$ in the following numerical examples. The SEM method is implemented by using Matlab on a MacBook Pro 2018 PC with 16 GB Memory and Intel Core i7 CPU. COMSOL was used for comparison verification on the same PC.
\subsection{Bloch periodic unit cell }
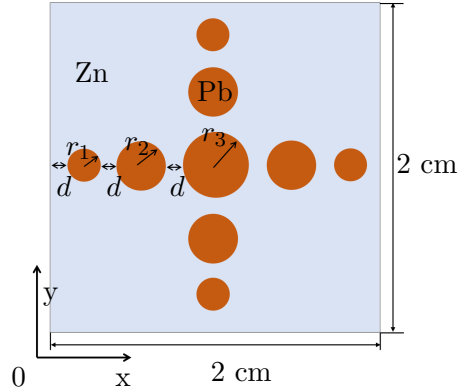
\begin{figure}[th] 
\centerline{\begin{overpic}[scale=0.25]
               {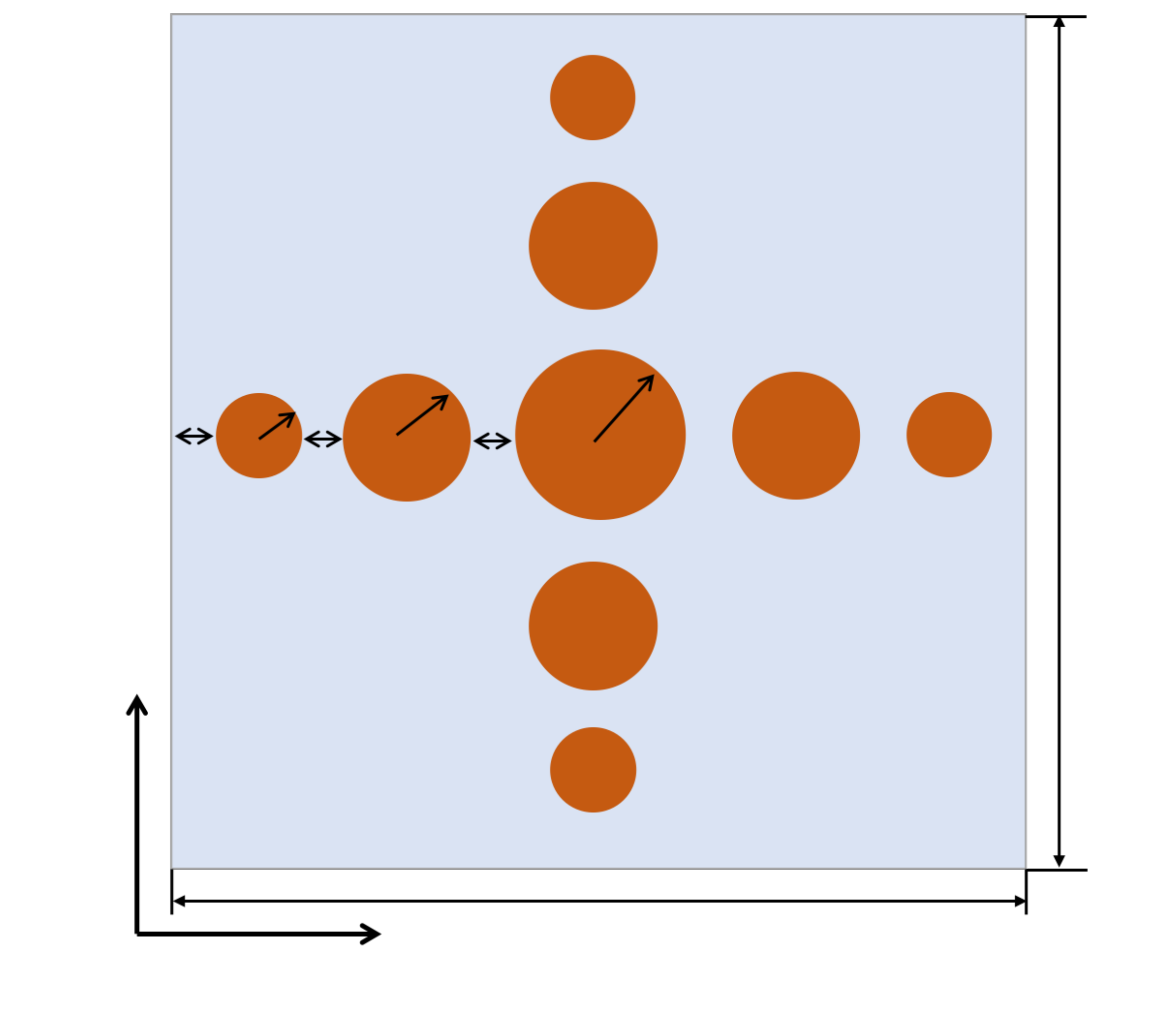}
               \put(20,68){Zn}
               \put(47,64){Pb}
               \put(18,52){$r_1$}
              \put(41,43){$d$}
              \put(27,43){$d$}
              \put(16,43){$d$}
                \put(31,53){$r_2$}
                \put(48,55){$r_3$}
                \put(29,1){x}
                \put(13,20){y}
                \put(50,2){2 cm}
                \put(91,48){2 cm}
                \put(6,1){0}
\end{overpic}}\vspace*{8pt}
\caption{schematic view of a unit cell with nine lead circles ($r_1=1$ mm, $r_2=1.5$ mm, $r_3=2$ mm). They are separated by the interval of $d=1$ mm and are embedded in the anisotropic zinc. The square outer boundaries are set as the Bloch periodic condition. }
\label{bpbc}
\end{figure}

In order to verify the accuracy  and convergence of the proposed SEM, we first consider a simple inhomogeneous anisotropic waveguide with the BPBC. The Bloch periodic unit cell has many applications in lithography and the design of elastic metasurfaces, which act as a plate-like waveguides connecting two elastic half-spaces \cite{Su2018}. The configuration of the unit cell is shown in Fig. \ref{bpbc}, where nine circular lead cores are embedded in the zinc square lattice. These circles with different radius are spaced one millimeter apart. The material properties are $\lambda_{\text{Pb}}=3.142\times10^{10}$ N/m$^2$, $\mu_{\text{Pb}}=5.986\times10^{9}$ N/m$^2$, $\rho_{\text{Pb}}=11340$ kg/m$^3$. The cladding is a transversely isotropic material with $c_{11}=16.5$ GPa, $c_{12}=3.1$ GPa, $c_{13}=5.0$ GPa, $c_{33}=6.2$ GPa, $c_{55}=3.96$ GPa and $\rho_{\text{Zn}}=2700 $kg/m$^3$. The frequency $f=5$ MHz and the unit cell is 2 cm containing multiple wavelengths, so that it is a large scale problem. 

\begin{figure}[th]   
\centerline{\begin{overpic}[width=12cm]
               {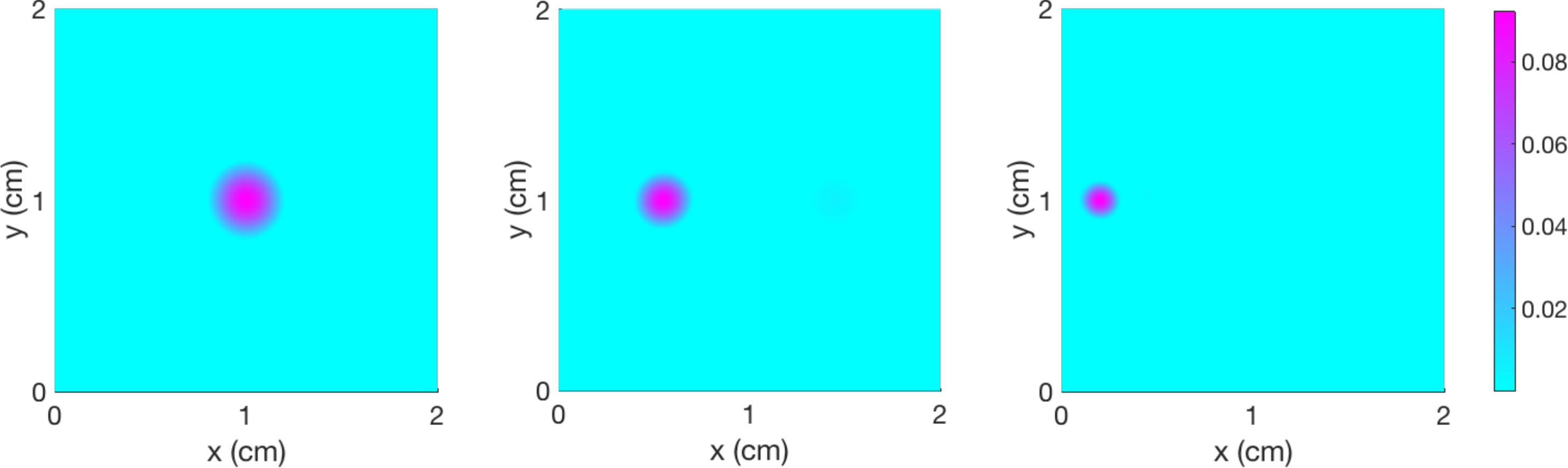}
               \put(5,25){(a)}
               \put(37,25){(b)}
               \put(70,25){(c)}
    \end{overpic}}\vspace*{8pt}
\caption{ Magnitude distributions of $\boldsymbol{\text{u}}(x,y)$ corresponding to $\mathrm{k}_{i,z}$ ($i=1, 7, 18$) of the BPBC waveguide shown in Figure \ref{bpbc}. (a)-(c) correspond to the 1st, 7th and 18th mode.}
\label{bpbc_u}
\end{figure}

\begin{figure}[th]   
\centerline{
\includegraphics[scale=0.32]{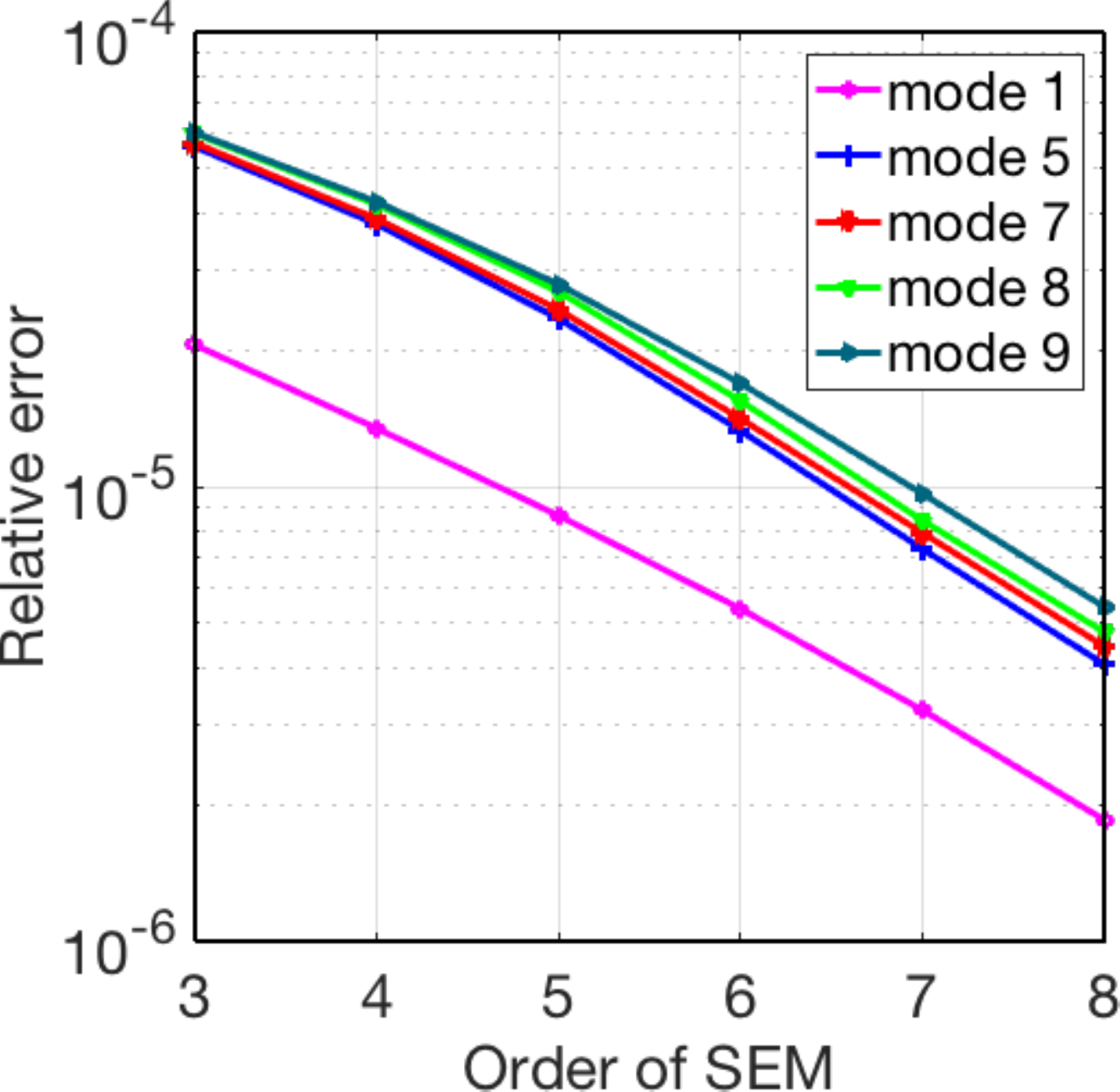}}\vspace*{8pt}
\caption{Relative errors of eigenmodes obtained by using the different order SEM for the BPBC waveguide shown in Figure \ref{bpbc}.}
\label{bpbc_re}
\end{figure}

The numerical results of k$_{i,z}^{N}$ obtained by the 5th-order SEM, the 5th-order FEM in COMSOL and the 10th-order SEM with an extremely fine mesh are shown in Table \ref{bpbc_eig} (the negligible imaginary part is not shown). They are denoted by SEM-K5, COMSOL-K5 and SEM-K10, respectively, and in view of the  maximum interpolation order of COMSOL is only 5, SEM-K10 is taken as the reference value. It is observed that SEM-K5 matches excellently with both COMSOL-K5 and SEM-K10. On the other hand, as illustrated in Table \ref{bpbc_com}, to achieve similar accuracy, COMSOL requires more 2.02 times DOFs, 2.35 times CPU time and 1.54 times memory than the SEM.  We can also see that when the numbers of element and DOF are taken to be similar, COMSOL is not as accurate as the SEM and requires a little more computational costs. Thus, the proposed SEM is more efficient than the FEM method, mainly because of the spectral accuracy (the exponential convergence) of the SEM shown in Fig. \ref{bpbc_re}. The magnitude distributions of $\boldsymbol{\text{u}}(x,y)$
for the 1st, 7th, 18th mode are displayed in Fig. \ref{bpbc_u}. All of them propagate in the lead core with different radii.
\begin{table}[th]%
\tbl{$\mathrm{k}_{z}$ (rad/s) of the elastic BPBC waveguide in Figure \ref{bpbc} obtained by the SEM and COMSOL.\label{bpbc_eig}}%
{\begin{tabular}{@{}cccc@{}}
\toprule
$i$& SEM-$\mathrm{k}_{i,z}^5$ & COMSOL-$\mathrm{k}_{i,z}^5$ & SEM-$\bar{\mathrm{k}}_{i,z}^{10}$\\
\hline
1-2& 4.3225E+04&4.3225E+04&4.3225E+04\\
3-10&4.3213E+04&4.3213E+04&4.3214E+04\\
11&4.3202E+04&4.3201E+04&4.3203E+04\\
12-13& 4.3201E+04&4.3200E+04&4.3201E+04\\
14&4.3199E+04&4.3199E+04&4.3199E+04\\
15-22&4.3180E+04&4.3181E+04&4.3183E+04\\
\midrule 
DOF&18543&37458&1305900\\
\bottomrule
\end{tabular}}
\end{table}

\begin{table}[th]%
\tbl{The comparison of the FEM and the SEM for the elastic BPBC waveguide in Figure \ref{bpbc}.\label{bpbc_com}}
{\begin{tabular}{@{}cccccc@{}}
\toprule
   & \# of Elements & DOF  &Error& Time (s)&Memory (GB)\\
\midrule 
FEM(N=5)& 507 & 37458 & 9.5E-6& 47 &2.08\\
FEM(N=5)& 264  &16488 & 1.1E-5& 29 &1.51\\
SEM(N=5)& 244  &18543 & 8.6E-6& 20 &1.35\\
\bottomrule
\end{tabular}}
\end{table}

\subsection{Resonant structure of an EMM}
\begin{figure}[th]   
\centerline{
\begin{overpic}[scale=0.22]
               {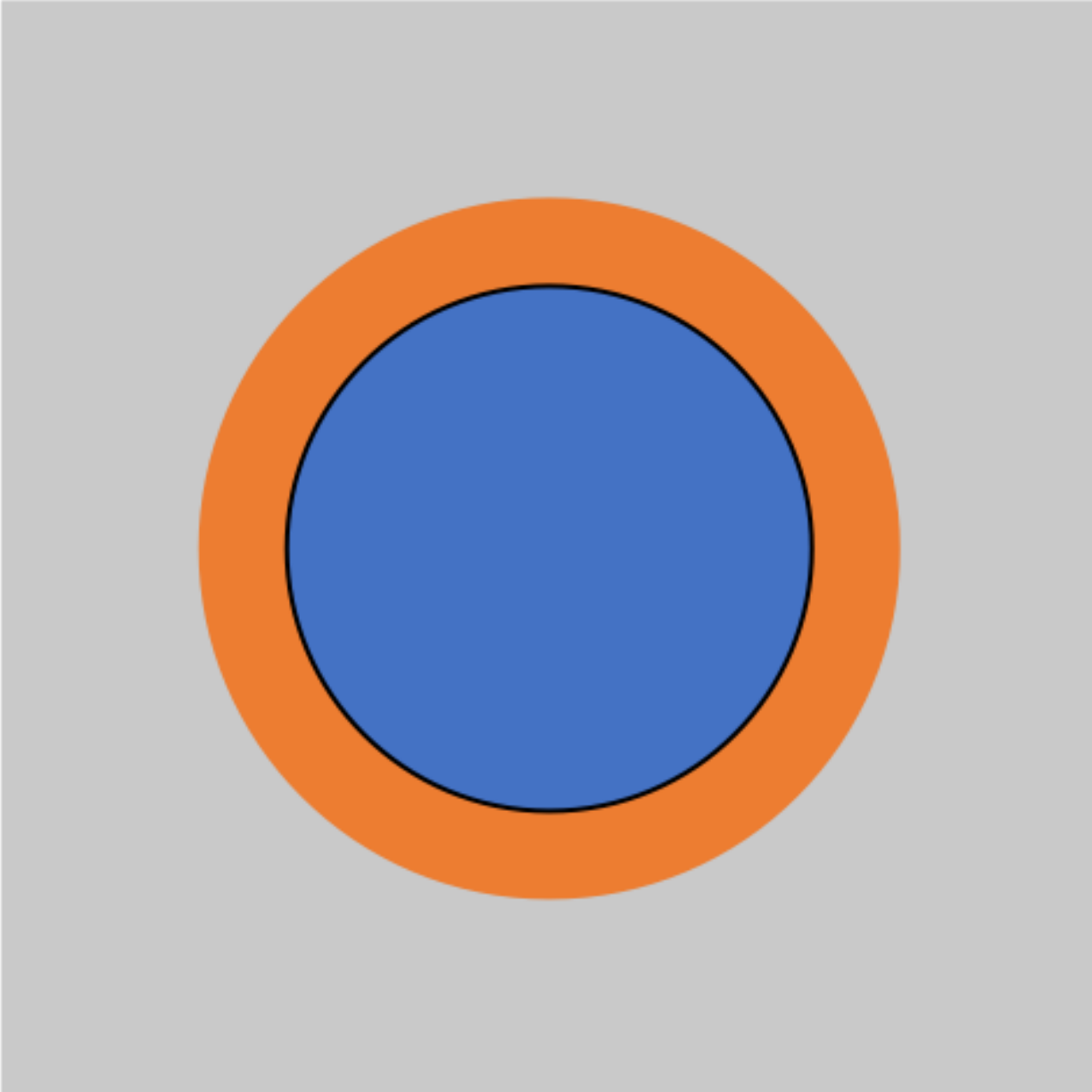}
  \put(38,6){foam}
  \put(34,74){rubber}
  \put(37,46){water}
\end{overpic}}\vspace*{8pt}
\caption{The cross section of the resonant structure for a left-hand material with a rubber coated water cylinder embedded in a foam host, with their material properties listed in Table \ref{para2}.}
\label{bloch_case_model}
\end{figure}

\begin{figure}[th]
\subfigure[$|{\textbf u}|$ for $\mathrm{k}_{1,z}$]{
\begin{minipage}[t]{0.2\textwidth}
\includegraphics[scale=0.3]{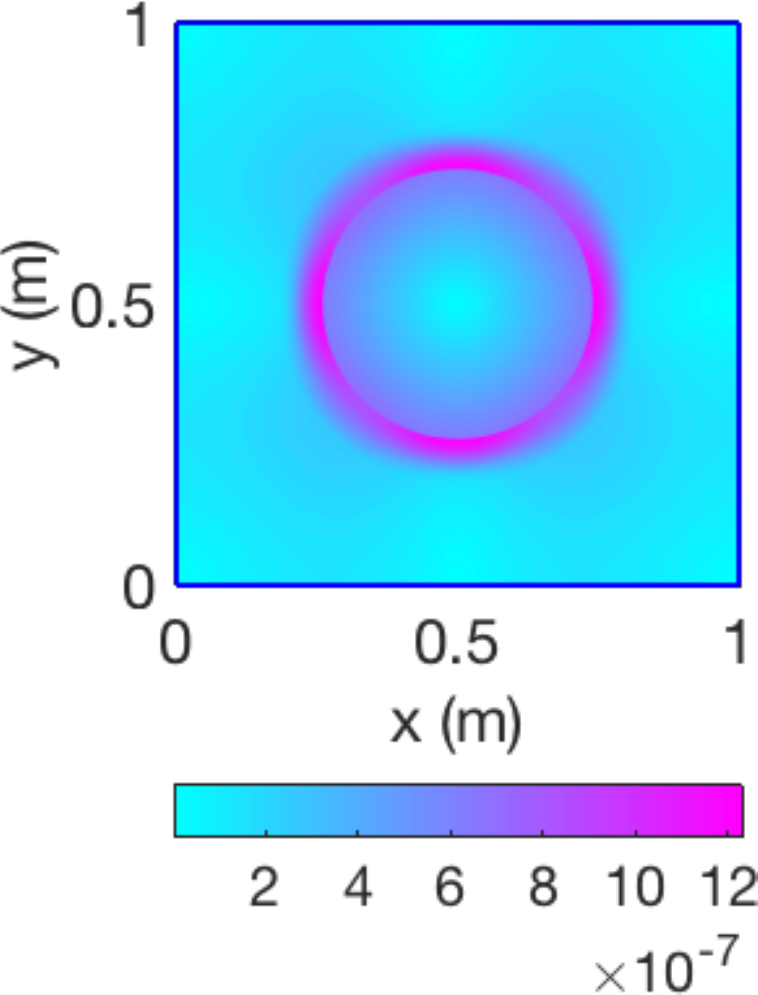}
\end{minipage}}
\subfigure[$|{\textbf u}|$ for $\mathrm{k}_{2,z}$]{\begin{minipage}[t]{0.2\textwidth}
\includegraphics[scale=0.3]{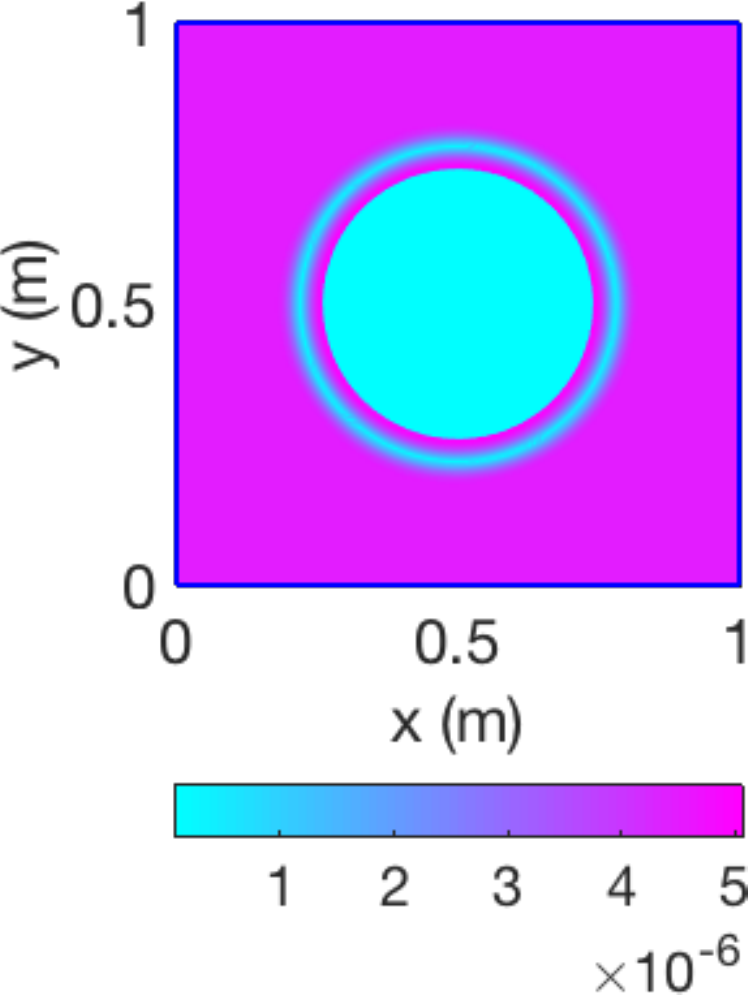}
\end{minipage}}
 \subfigure[3D vector of ${\textbf u}$ for $\mathrm{k}_{1,z}$]{\begin{minipage}[t]{0.3\textwidth}
\includegraphics[scale=0.3]{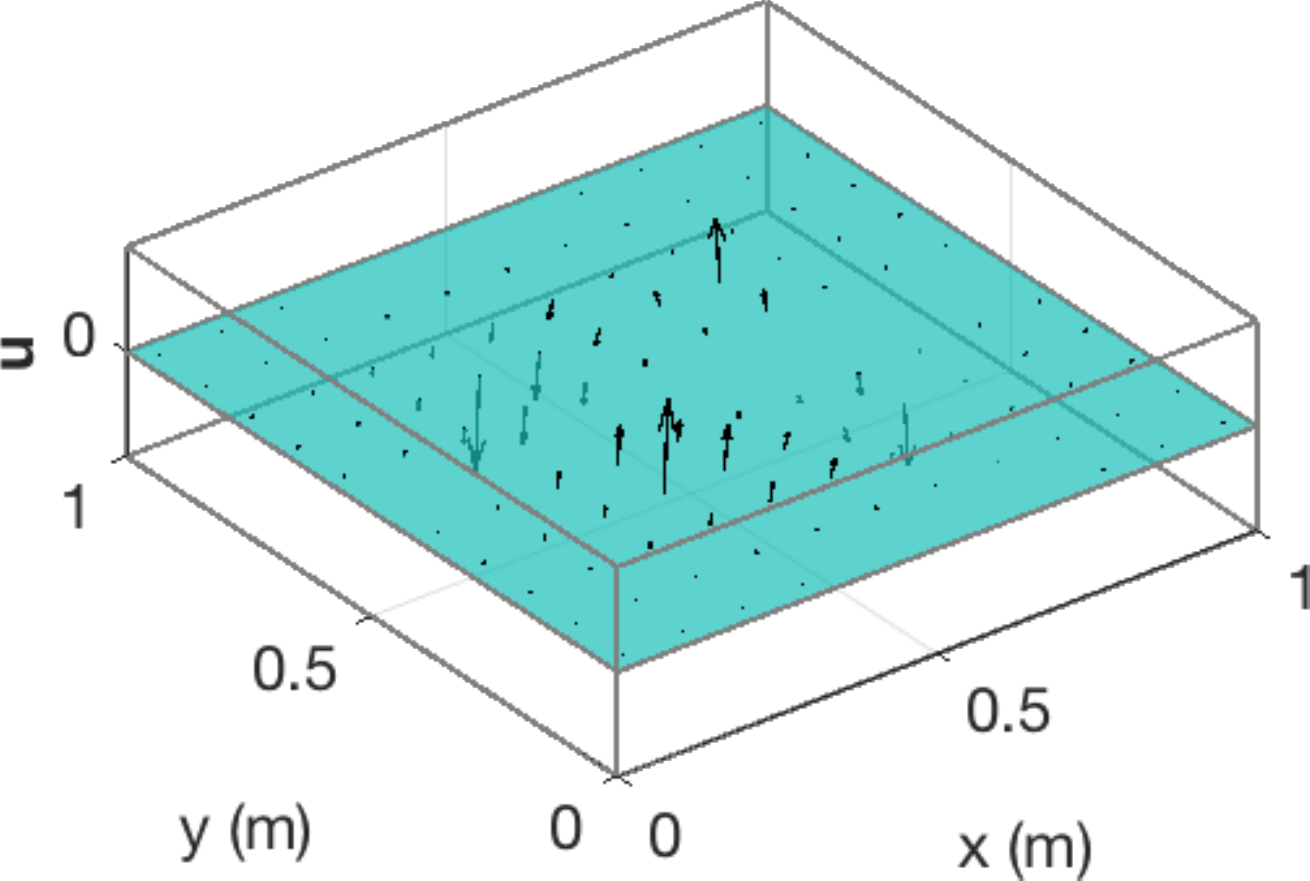}
\end{minipage}}
\subfigure[3D vector of ${\textbf u}$ for $\mathrm{k}_{2,z}$]
{\begin{minipage}[t]{0.2\textwidth}
\includegraphics[scale=0.3]{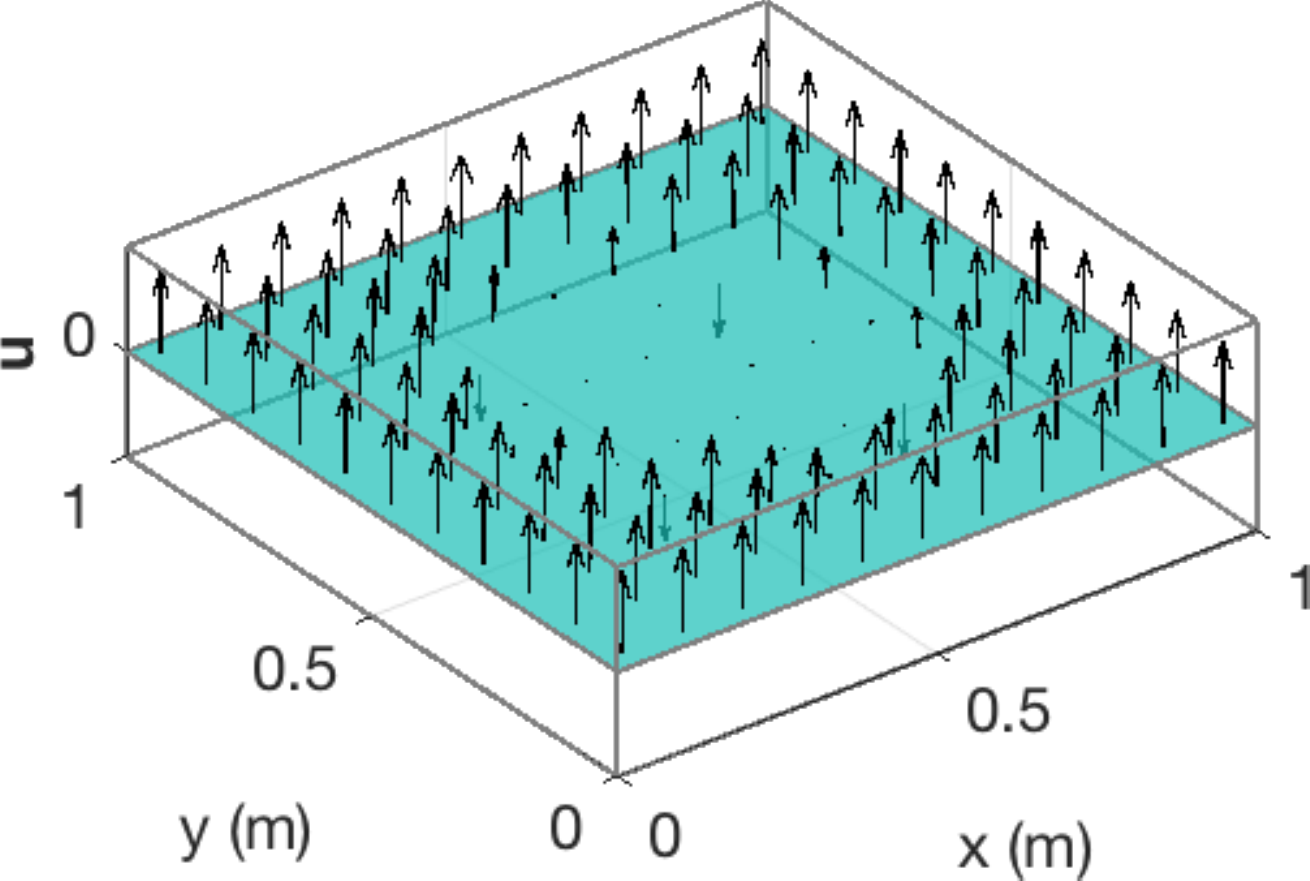}
\end{minipage}}\vspace*{8pt}
\caption{The distributions of ${\textbf u}$ correspond to the first two modes in the elastic resonant structure waveguide in Figure \ref{bloch_case_model}. The first mode in (a) and (c) exhibits a quadrupolar resonance in the rubber region. The second mode in (b) and (d) shows a total reflection at the interface between the rubber region and the water region.}
\label{Resonant Structure}
\end{figure}

In order to verify that our SEM solver is accurate and efficient for the inhomogeneous solid-fluid coupling BPBC waveguide, we first consider a resonant structure in the building block of a left-handed material (LHM) proposed in \cite{WuElastic}. This kind of resonant structure will bring negative elastic parameters within a certain frequency range and the cross section of the unit cell, a rubber coated water cylinder embedded in a foam host, is shown in Fig. \ref{bloch_case_model}. The lattice constant is $a$ and the radius of the rubber and water is $0.32a$ and $0.24a$, respectively.  When we set $a=1$ m, the corresponding frequency is chosen as 34.887 Hz referring to \cite{WuElastic}. In addition, the material parameters are listed in Table \ref{para2} and the BPBC is used in the example, $(\theta,\phi)=(0,0)$ and $k=\omega/v_s$, $v_s$ is the velocity of  S wave in the foam. Table \ref{RS eig} shows that the numerical solutions of the inhomogeneous isotropic BPBC waveguide obtained by the SEM and COMSOL agrees well. On the other hand, as illustrated in Table \ref{rs_com}, the proposed SEM is more efficient than the FEM in terms of the DOF and memory. Moreover, from the subgraph (a) and (c) of Fig. \ref{Resonant Structure}, we can see that there is a quadrupolar resonance in the rubber region for the first mode produced by the P wave, due to the much smaller $v_p$ of the rubber than those in the background foam and the water core. Besides, as shown in the subgraph (b) and (d) of Fig. \ref{Resonant Structure}, a total reflection occurs at the boundary between the rubber region and the water core for the second mode, because of the much larger $v_p$ of the water in Table \ref{para2}. 

\begin{table}[th]%
\tbl{Parameters for the Resonant Structures in Figure \ref{bloch_case_model}.\label{para2}}%
{\begin{tabular}{@{}lccccc@{}}
\toprule
          & $\lambda$ (N/m$^2$) &$\mu$ (N/m$^2$)& $\rho$ (kg/m$^3$) & $v_p$ (m/s)&$v_s$ (m/s)\\
\midrule 
    foam&$6\times10^6$&$3\times10^6$&115&323&161.5\\
  rubber&$6\times10^5$&$4\times10^4$&$1300$&22.8&5.5\\
 water&$2.25\times10^9$ &0&1000&1500&0\\
\bottomrule
\end{tabular}}
\end{table}

\begin{table}[th]%
\tbl{The first two values of $\mathrm{k}_{z}$ (rad/s) of the elastic resonant structure waveguide in Figure \ref{bloch_case_model} obtained by the SEM and COMSOL.\label{RS eig}}%
{\begin{tabular}{@{}cccc@{}}
\toprule
   $i$& SEM~-~$\mathrm{k}_{i,z}^5$ & COMSOL~-~$\mathrm{k}_{i,z}^5$ & SEM~-~$\bar{\mathrm{k}}_{i,z}^{10}$\\
          \hline
1& 2.85094& 2.85087&2.85093\\
2&0.73831& 0.73831&0.73831\\
\midrule 
DOF&18523&20401&444278\\
\bottomrule
\end{tabular}}
\end{table}

\begin{table}[th]%
\tbl{The comparison of FEM and SEM for the elastic resonant structure waveguide in Figure \ref{bloch_case_model}.\label{rs_com}}%
{\begin{tabular}{@{}ccccc@{}}
\toprule
   & \# of Elements & DOF  &Error& Memory (GB)\\
\midrule 
FEM(N=5)& 308 & 20401 & 1.8E-5& 1.33 \\
FEM(N=5)& 408  &25601 & 8.0E-6& 1.35 \\
SEM(N=5)& 297  &18523 & 5.6E-6&0.88 \\
\bottomrule
\end{tabular}}
\end{table}
 
\subsection{Optical fiber model}
\begin{figure}[th]
\centerline{\begin{overpic}[width=4cm]%
               {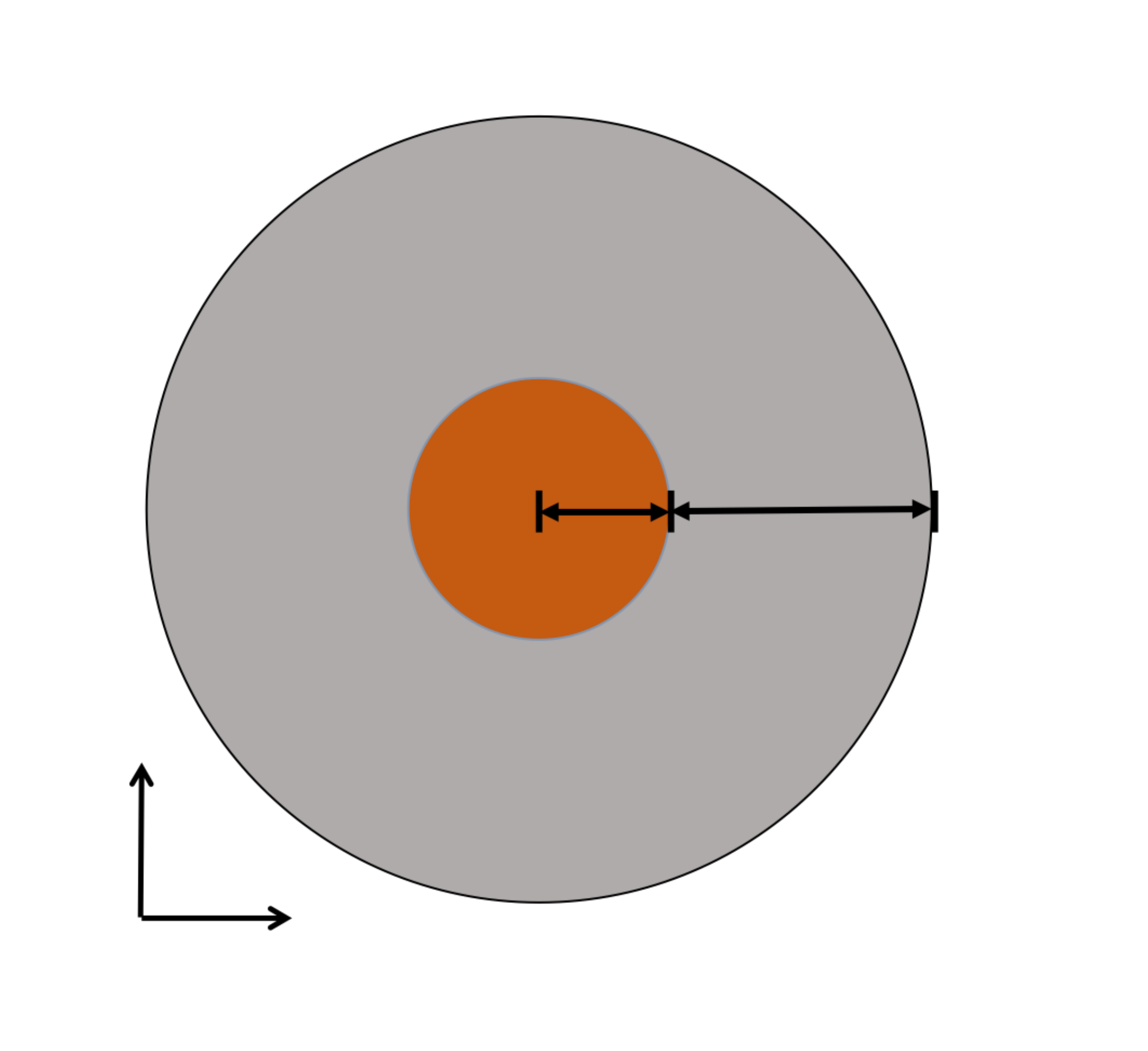}
  \put(41,50){core}
  \put(33,68){cladding}
  \put(50,42){a}
  \put(66,42){2a}
  \put(26,11){$x$}
  \put(13,24){$y$}
\end{overpic}}\vspace*{8pt}
\caption{The cross section of the optical fiber (a=4.1 $\mu$m), with an unbounded cladding truncated by an ABC at $r=3a$.  The core and cladding materials can take the combination of materials listed in Table \ref{para1}.}
\label{optical_fiber_model}
\end{figure}
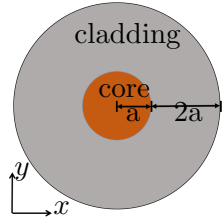

Next, to verify the ABC formulation for an open (unbounded) inhomogeneous isotropic waveguide, we consider the optical fiber. It is a common optical waveguide consisting of the cladding and the fiber core, and its elastic waveguide properties are of significant interest  \cite{zou2006two,wolff2015stimulated}. The cross section of the optical fiber is shown in Fig. \ref{optical_fiber_model}, which consists of the core and the cladding. The radius of the core and the cladding is $a=4.1$ $\mu$m and $3a=12.3$ $\mu$m, respectively; the cladding is pure SiO$_2$ and
the core is  filled with one of the three different materials as shown in Table \ref{para1}. The SEM is employed to simulate the elastic waveguide properties of this optical fiber. Besides, to verify the accuracy and effectiveness of the SEM for solving the solid-fluid system, a fluid cladding is also considered. The material parameters are included in Table \ref{para1}. In order to simulate the unbounded waveguide structure, the ABC is used to truncate the cladding so that the simulated structure mimics an infinite cladding region.

\begin{table}[th]%
\tbl{Parameters for the Cores and Claddings of the Optical Fiber in Figure \ref{optical_fiber_model}.\label{para1}}%
{\begin{tabular} {@{}lccccc@{}}
\toprule
    & $v_p$ (m/s)& $v_s$ (m/s)& $\rho$ (kg/m$^3$)&$\lambda$ (N/m$^2$)& $\mu$ (N/m$^2$)\\
\midrule 
  core 1 (impure SiO$_2$)&5794.626 &3644.85&2291.25& 1.6057$\times10^{10}$&30.44$\times10^9$\\
 core 2 (EMM)&829.89 &532.53j& -1481&-1.86$\times10^9$&0.42$\times10^9$\\
  core 3 (normal)&1202 &532.53 &1481&1.30$\times10^9$ &0.42$\times10^9$\\
 cladding 1 (SiO$_2$)&5970& 3760&2201 &1.6212$\times10^{10}$&31.13$\times10^9$\\
 cladding 2 (water)&1500 & 0 & 1000 &2.25$\times10^9$ &0\\
\bottomrule
\end{tabular}}
\end{table}

\begin{enumerate}[(a)]
\item Normal Elastic Materials\\
   First, to verify the accuracy of the SEM solver for the inhomogeneous open waveguide problems, we conduct a numerical experiment on the actual optical quartz fiber model consisting of cladding 1 and core 1, and the frequency is chosen as 3 GHz as in realistic application \cite{AamirElimination}. The agreement among the three results in Table \ref{of eig} verifies the accuracy of our scheme. Besides, we observe that the real part of the higher-order mode gradually decreases while the imaginary part falling into different orders of magnitude gradually increases. The phenomenon indicates that the energy loss of the higher-order mode gradually increases. The relative errors and computational costs of the SEM and the FEM are illustrated in Table \ref{op_com}. For a similar mesh, the proposed 3rd-order SEM and 4th-order FEM can achieve similar accuracy (3E-6). The corresponding memory used by FEM is more than SEM,  illustrating the proposed SEM is more efficient than FEM. Moreover, it can be observed that the 6th-order SEM can achieve higher accuracy (3E-7) with less memory, due to the spectral accuracy of the SEM solver. Incidentally, no impurity is present in our contour maps in Fig. \ref{optical_fiber_1}, which indirectly indicates that  no spurious modes exist in our method as discussed in \cite{AamirElimination}. In addition, there are convincing explanations for the spurious modes. In general, spurious modes are obtained in the following two cases. One is that basis functions cannot describe the physical properties of solutions. In this manuscript, the GLL polynomials are employed to construct the basis functions which are obviously continuous at the interpolation points. The other one is that a discrete space cannot compactly approximate the solution space $H^1(\Gamma)$. For our method, the discrete space $Q^{N,h}=\text{span}\{ \phi_1, \phi_2, \cdot, \phi_{N_{sj}}\}$ is used to approximate the solution space $ H^1(\Gamma)$, so that it is compact. In conclusion, there are no spurious modes in our method. Moreover, waves are well absorbed at the outer absorbing boundary.
    
\begin{table}[th]%
\tbl{$\mathrm{k}_{z}$ (Mrad/s) of the elastic fiber-optics waveguide in Figure \ref{optical_fiber_model} obtained by the SEM and COMSOL for Core 1 and Cladding 1 listed in Table \ref{para1}.\label{of eig}}%
{\begin{tabular} {@{}cccc@{}}
\toprule
$i$& SEM~~-$\mathrm{k}_{i,z}^5$ & COMSOL~-~$\mathrm{k}_{i,z}^5$ & SEM~-~$\bar{\mathrm{k}}_{i,z}^{10}$\\
\hline
1-2& 5.1478-3.8741E-11j&5.1478-3.8627E-11j&5.14780-3.8606E-11j\\
3&5.1127-7.1091E-10j&5.1127-7.1090E-10j&5.1127-7.1090E-10j\\
4-5&5.1119-9.6837E-10j&5.1119-9.6650E-10j&5.1119-9.6616E-10j\\
6& 5.1116-1.2307E-09j&5.1116-1.2265E-09j&5.1116-1.2256E-09j\\
7-8&5.0670-3.7985E-08j&5.0670-3.7920E-08j&5.0670-3.7907E-08j\\
\midrule 
DOF&33648&33447&249303\\
\bottomrule
\end{tabular}}
\end{table}

\begin{table}[th]%
\tbl{The comparison of FEM and SEM for the elastic fiber-optics waveguide in Figure \ref{optical_fiber_model}  with Core 1 and Cladding 1 listed in Table \ref{para1}.\label{op_com}}%
{\begin{tabular} {@{}ccccc@{}}
\toprule
    &\# of Elements & DOF  &Error &Memory (GB)\\
  \midrule
FEM(N=3)& 112 & 3207 & 3.8E-5&1.37 \\
FEM(N=4)& 112  &5619 & 1.9E-6&1.42 \\
SEM(N=3)& 107  &2982 & 3.7E-6&0.79 \\
SEM(N=6)& 107  &11739& 3.0E-7&1.04\\
\bottomrule
\end{tabular}}
\end{table}
 
\begin{figure}[th]
\subfigure[Mode for $\mathrm{k}_{1,z}$]{\begin{minipage}[t]{0.55\textwidth}
\includegraphics[scale=0.16]{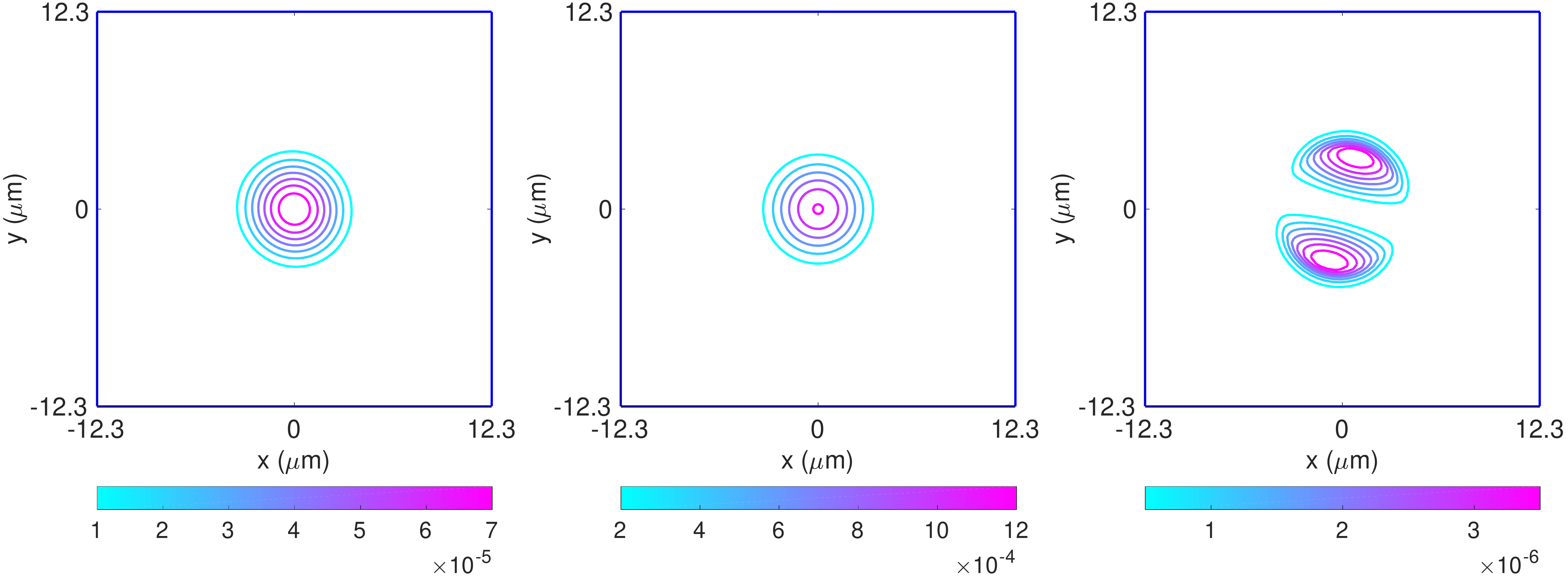}
\end{minipage}}
\subfigure[Mode for $\mathrm{k}_{2,z}$]
{\begin{minipage}[t]{0.4\textwidth}
\includegraphics[scale=0.16]{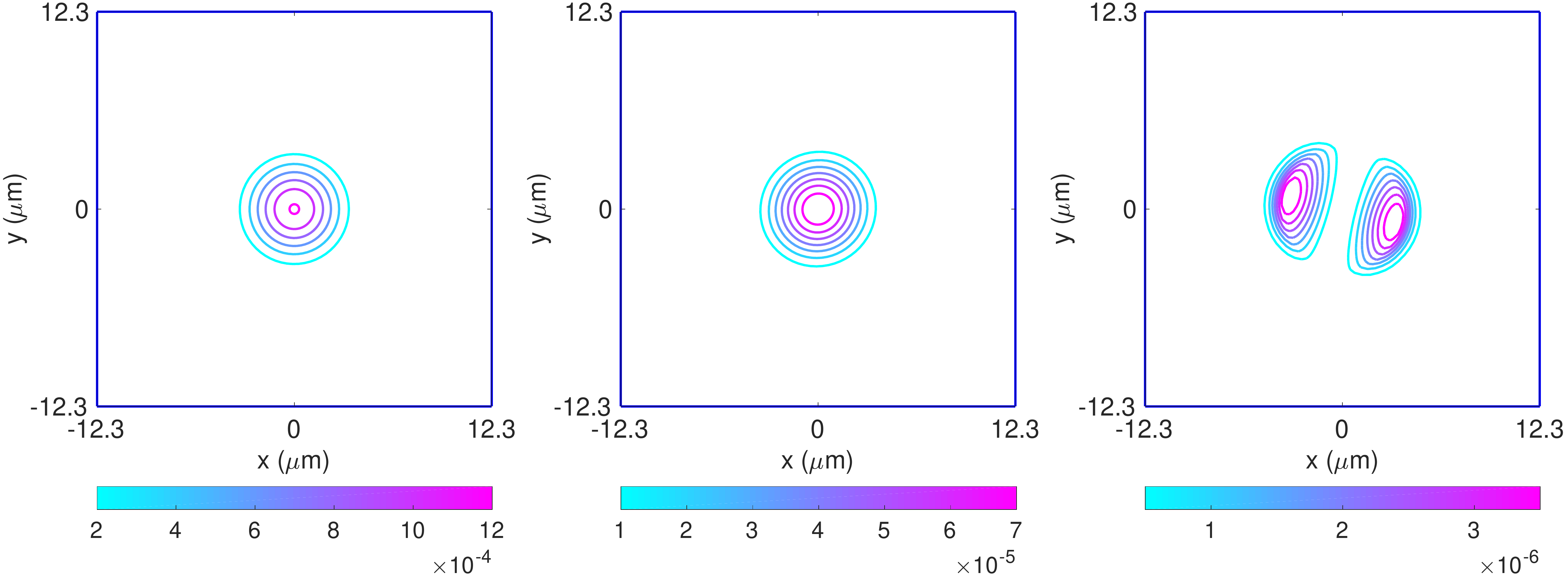}
\end{minipage}}

\subfigure[Mode for $\mathrm{k}_{3,z}$]{\begin{minipage}[t]{0.55\textwidth}
\includegraphics[scale=0.16]{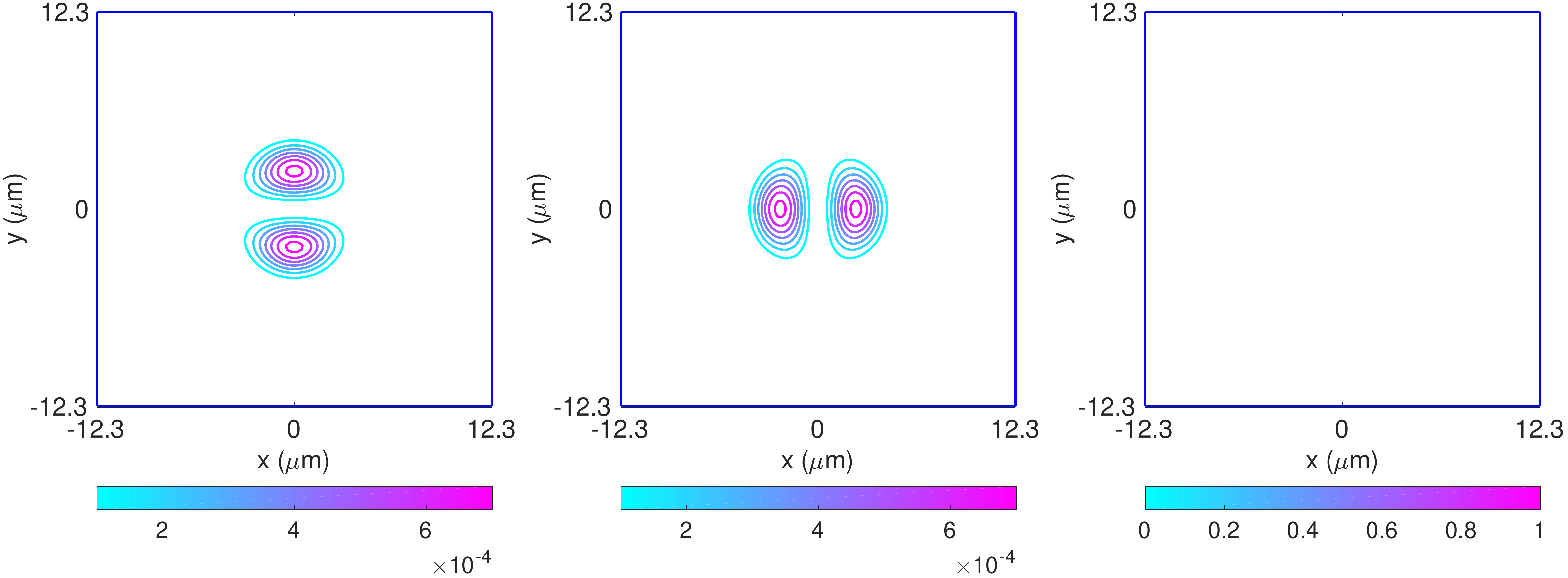}
\end{minipage}}
\subfigure[Mode for $\mathrm{k}_{4,z}$]
{\begin{minipage}[t]{0.4\textwidth}
\includegraphics[scale=0.16]{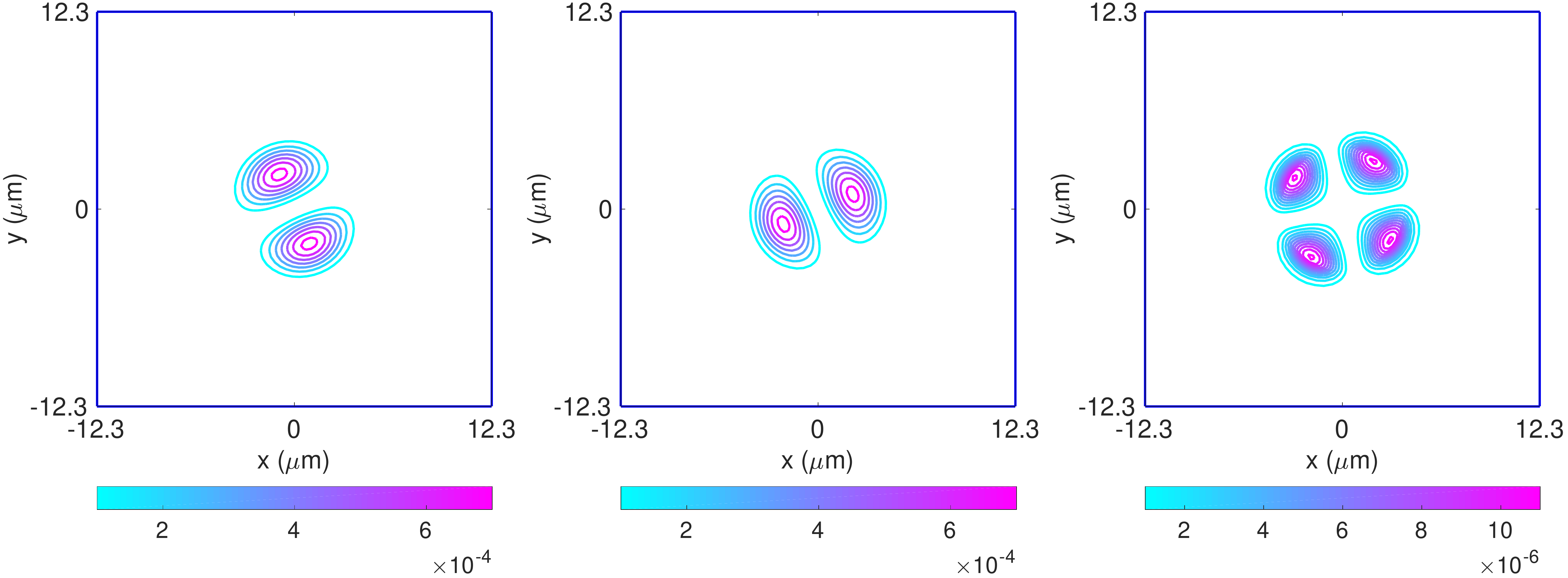}
\end{minipage}}\vspace*{8pt}
\caption{Contour maps of $u_x$, $u_y$, $u_z$ in the open fiber-optics waveguide problem in Figure \ref{optical_fiber_model} with an impure SiO$_2$ core 1. (a)-(d) correspond to the first to the fourth mode. No impurity shown in the contour maps indicate that no spurious modes exist and waves are well absorbed at the outer absorbing boundary.}
\label{optical_fiber_1}
\end{figure}

\item Double Negative Index Elastic Metamatrial (EMM) Core\\
Second, for the same size model, we now consider the effects of the EMM core with a negative index. We design an example on a simultaneously negative mass density and bulk modulus EMM core 2 constructed by reference \cite{LiuAn}, embedded in Cladding 1 in Table \ref{para1}. The frequency is chosen as 0.3 GHz. Through calculating the velocities of P-wave and S-wave respectively shown in Table \ref{para1}, we find the velocity of S-wave is an imaginary number, thus the S-wave is forbidden in this material. Again, k$_{i,z}^{N}$ ($N$=5,10) obtained by the two methods are shown in Table \ref{of eig2}. It is observed that the SEM solution matches excellently with the reference results and the COMSOL's results, verifying that our scheme is suitable for the negative index materials. Meanwhile, as illustrated in Table \ref{op_emm1_com}, DOF and the memory used by COMSOL ($N$=5) is 2 and {\color{re1}{1.5}} times more than SEM  ($N$=5) to achieve the similar accuracy (8E-7). Evidently, it shows the high computation efficiency  of the SEM. Furthermore, the magnitude distributions of $\boldsymbol{\text{u}}(x,y)$ corresponding to $\mathrm{k}_{i,z}$ $(i= 1,\cdots,6)$ with EMM core 2 are plotted in Fig. \ref{optical_fiber_2}. On the other hand, instead of core 2, we conduct another experiment on core 3, whose density and bulk modulus are positive. The agreement in Table \ref{op_emm1_com} verifies the accuracy of the results. Fig. \ref{optical_fiber_3} plots the distribution of  $\boldsymbol{\text{u}}$ corresponding to $\mathrm{k}_{i,z}$ $(i= 1,2,3)$ with core 3. In contrast to the previous configuration, we observe in Fig. \ref{optical_fiber_2} that these modes in the waveguide of EMM core 2 propagate only at the interface between the core and cladding because of the presence of the negative index material.

\begin{table}[th]
\tbl{$\mathrm{k}_{z}$ (Mrad/s) of the elastic waveguide with either EMM  Core 2 or Core 3 and Cladding 1 in Figure \ref{optical_fiber_model} obtained by the SEM and COMSOL.\label{of eig2}}
{\begin{tabular}{@{}lccccc@{}}
\toprule
&\multicolumn{3}{c}{EMM core 2}& \multicolumn{2}{c}{core 3}\\
    \cmidrule(lr){2-4} \cmidrule(lr){5-6}
$i$& SEM-$\mathrm{k}_{i,z}^5$ & COMSOL-$\mathrm{k}_{i,z}^5$ & SEM-$\bar{\mathrm{k}}_{i,z}^{10}$ & SEM-$\mathrm{k}_{i,z}^5$&COMSOL-$\mathrm{k}_{i,z}^5$\\
 \midrule
1&-3.0272216&-3.0272236&-3.0272209&3.4871638&3.4871639\\
2&-3.0139166&-3.0139190&-3.0139156&3.4142176&3.4142180\\
3&-2.9737232&-2.9737341&-2.9737218&3.4041896&3.4041900\\
4&-2.9056978&-2.9057024&-2.9056937&3.3968465&3.3968469\\
5&-2.8080154&-2.8080351&-2.8080125&3.3021688&3.3021695\\
\bottomrule
DOF&11433&24138&189963&11433&24138\\
\bottomrule
\end{tabular}}
\end{table}

\begin{table}[th]%
\tbl{The comparison of FEM and SEM with EMM Core 2 in Table \ref{of eig2}.\label{op_emm1_com}}%
{\begin{tabular} {@{}cccccc@{}}
\toprule
 & \# of Elements & DOF  &Error& Memory (GB) \\
\midrule 
FEM(N=5)& 157 & 12018 & 5.64E-6& 1.39\\
FEM(N=5)& 317  &24138 & 8.93E-7& 1.62\\
SEM(N=5)& 150  &11433 & 8.16E-7&1.19 \\
\bottomrule
\end{tabular}}
\end{table}

\begin{figure}[th]
\centerline{\begin{overpic}[width=10cm]%
               {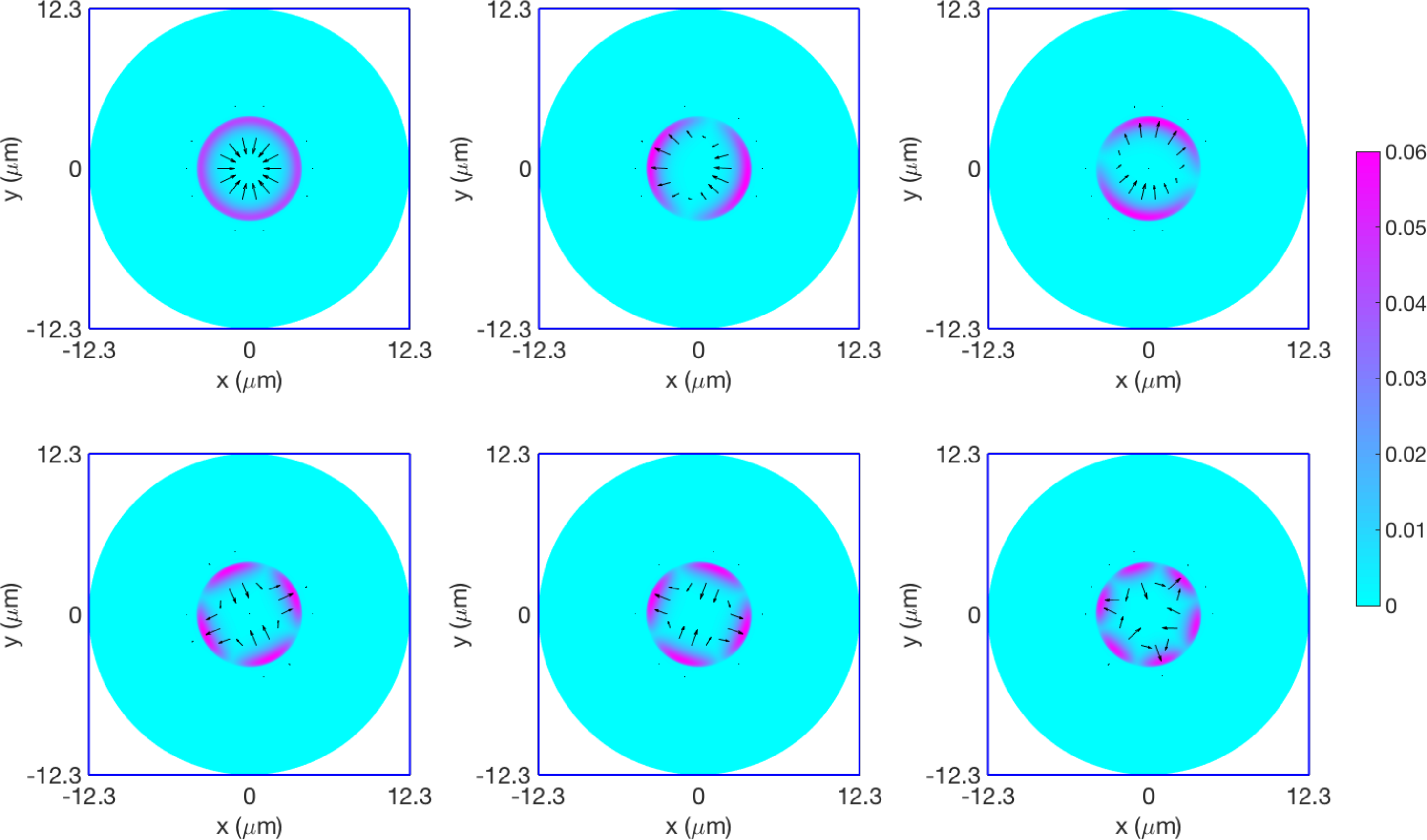}
               \put(6,55){(a)}
               \put(38,55){(b)}
	       \put(70,55){(c)}
	       \put(6,24){(d)}
               \put(38,24){(e)}
	       \put(70,24){(f)}
\end{overpic}}\vspace*{8pt}
\caption{ Magnitude distributions of $\boldsymbol{\text{u}}(x,y)$ for eignmodes corresponding to $\mathrm{k}_{i,z}$ ($i=1, \cdots , 6$) obtained in the open fiber-optics waveguide problem with EMM core 2 in Cladding 1 in Figure \ref{optical_fiber_model}. (a)-(f) correspond to the first to the sixth mode. All of them propagate only at the interface between the core and cladding because of the presence of the negative index material, different from the normal material Core 3 in Fig. \ref{optical_fiber_3}.}
\label{optical_fiber_2}
\end{figure}

\begin{figure}[th]
\centerline{\begin{overpic}[width=10cm]%
               {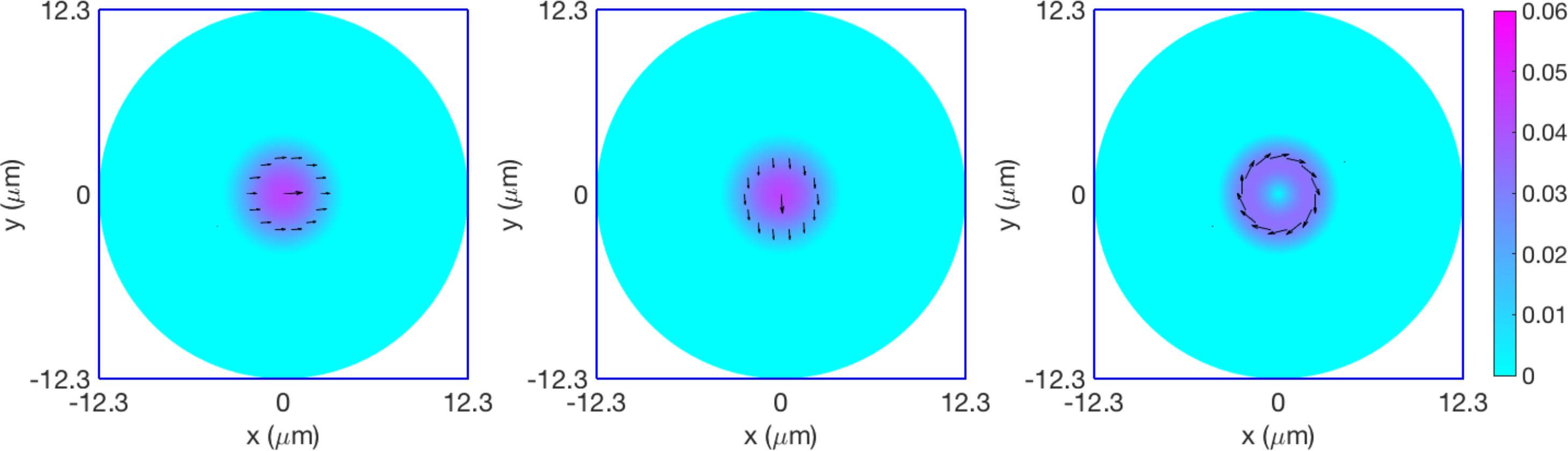}
               \put(7,25){(a)}
               \put(39,25){(b)}
	       \put(71,25){(c)}
\end{overpic}}\vspace*{8pt}
\caption{Magnitude distributions of $\boldsymbol{\text{u}}(x,y)$ for eignmodes corresponding to $\mathrm{k}_{i,z}$ ($i=1,2,3$) obtained in the open fiber-optics  waveguide problem in Figure \ref{optical_fiber_model} with the normal material core 3 in Cladding 1. (a)-(c) correspond to the first to the third mode. In the contrast to the EMM Core 2 in Figure \ref{optical_fiber_2}, the fundamental mode found in the normal material open fiber-optics waveguide is concentrated in the whole core region.}
\label{optical_fiber_3}
\end{figure}
Furthermore, we notice that one propagation mode exists under low frequencies ($fd/c\in[6.5\times10^{-6},9\times10^{-3}]$, $d=2a$)  as shown in Fig. \ref{dispersion1}. Within this frequency range, different from normal elastic materials, the increase of the frequency does not alter the distribution interval of the real part of $\mathrm{k}_z$ (the phase constant $\beta_z=-5.23\times10^5$) with the negligible imaginary part (the attenuation constant $\alpha_z$) on the basis of the positive $P_z$. Through the observation in Fig. \ref{lowfrequency}, we can find the propagation mode is caused by the P wave and concentrated in the core.  In addition, in order to explain the existence of this mode, the phase velocity $v_p=\omega/\beta_z$ is shown in Fig. \ref{dispersion1} (b). It can be found that this mode exhibits backward wave propagation in the cross section, which is defined as the phase velocity direction ($-\hat{z}$) antiparallel to the Poynting vector ($+\hat{z}$), caused by the negative-index materials \cite{bramhavar2011negative}. Hence, different from normal elastic materials, the application of EMMs will bring some special eigenmodes in the elastic waveguide.
\begin{figure}[th]
\begin{minipage}[h]{0.45\linewidth}
\centerline{\begin{overpic}[width=5cm]{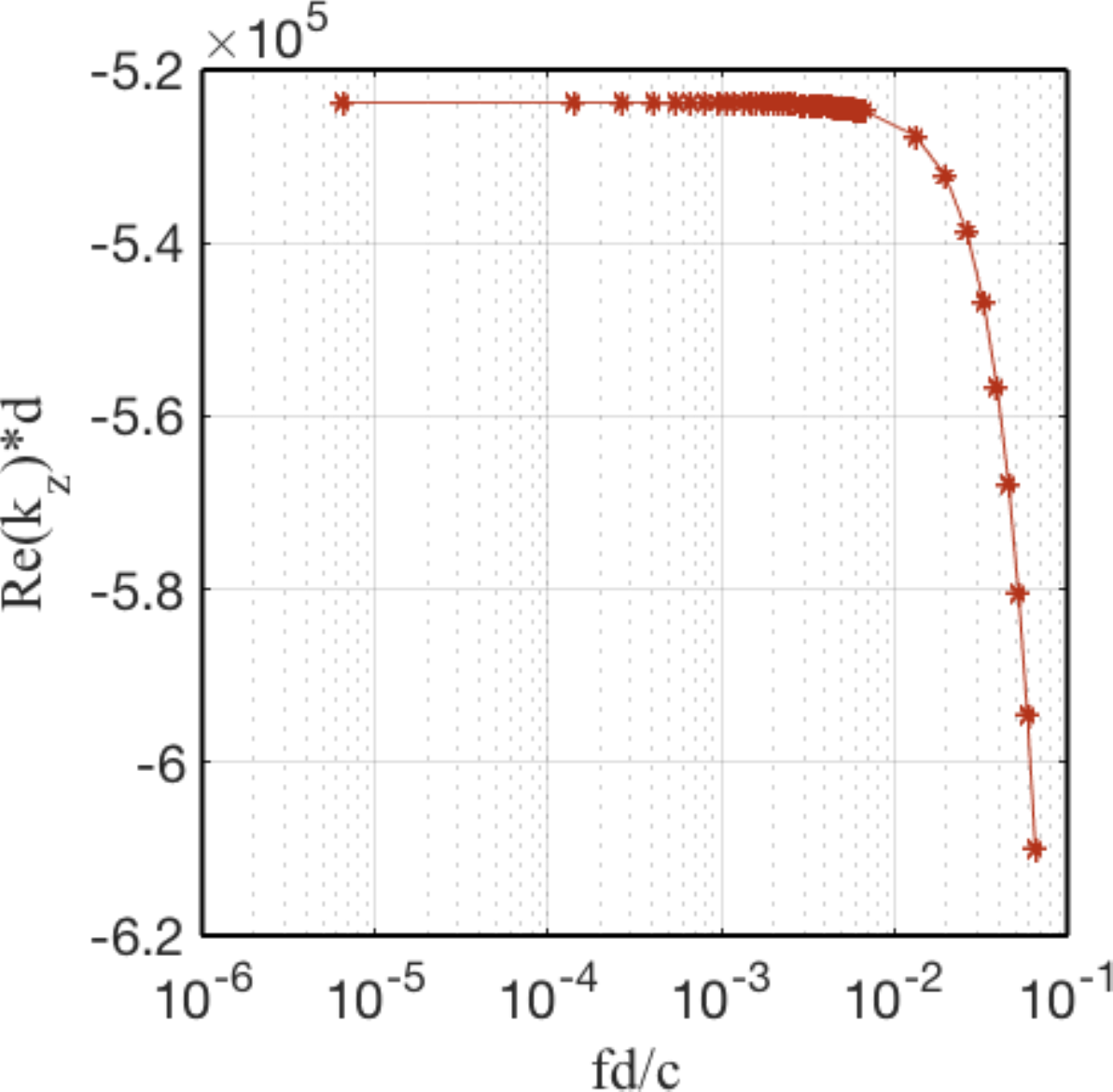}\put(25,75){(a)}\end{overpic}}
\end{minipage}
\begin{minipage}[h]{0.5\linewidth}
\centerline{\begin{overpic}[width=5cm]{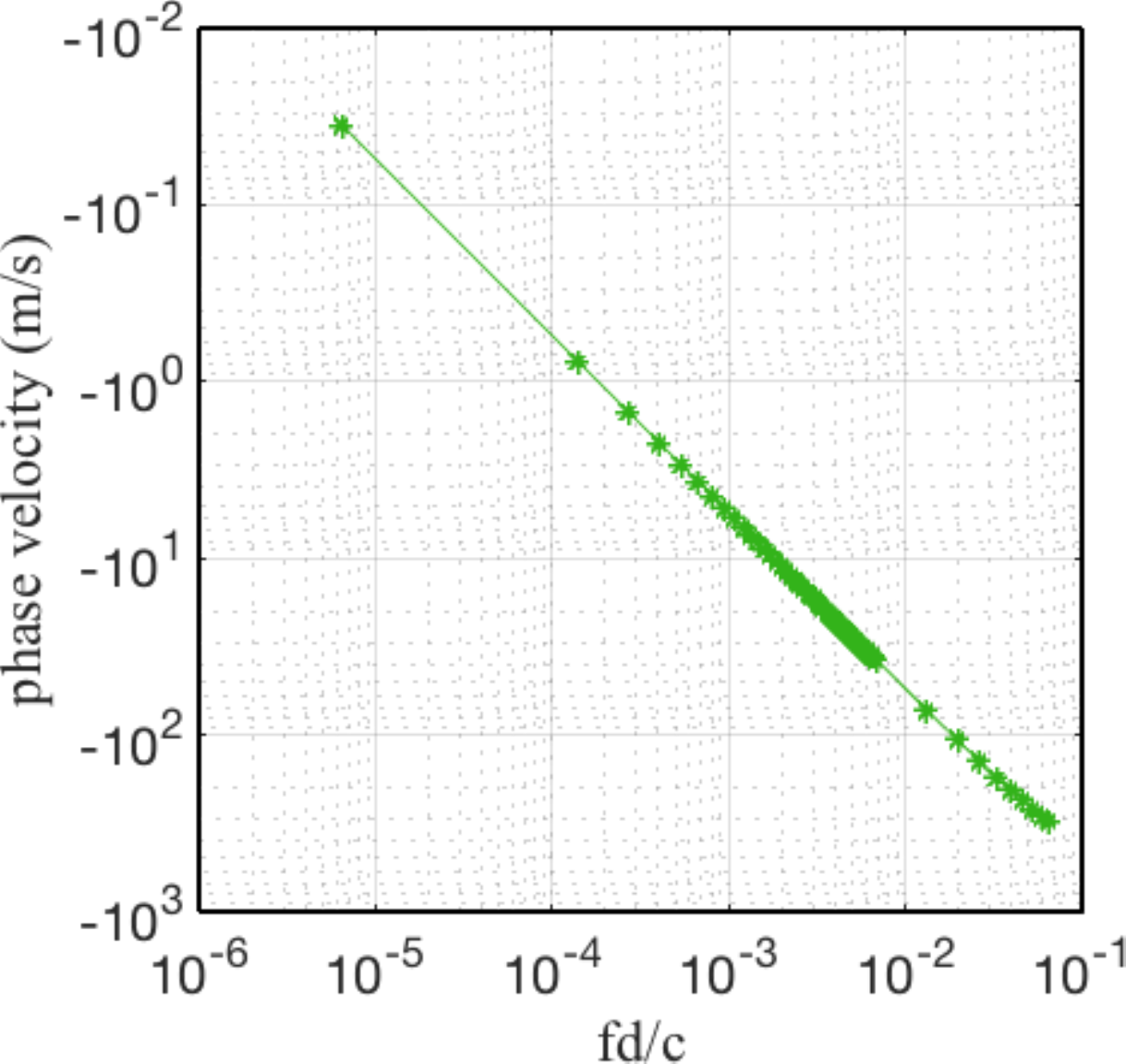}\put(25,75){(b)}\end{overpic}}
\end{minipage}\vspace*{8pt}
\caption{The dispersion curves versus with frequency ($d=2a$, $c=v_s$ of Cladding 1) for the fundamental mode in the waveguide with EMM Core 2 and Cladding 1 in Figure \ref{optical_fiber_model}. (a) The real part of of $k_z$. Within the low frequency  $fd/c\in[6.5\times10^{-6},9\times10^{-3}]$, the real part of $k_z$ does not vary with the frequency. (b) The phase velocity. The negative value means the direction of the phase velocity is $-\hat{z}$, antiparallel to the $+\hat{z}$. Thus, the backward wave propagation phenomenon is found in this mode.}
\label{dispersion1}
\end{figure}

\begin{figure}[th]
\subfigure[$|{\textbf u}|$]{\begin{minipage}[h]{0.3\linewidth}
\centerline{\includegraphics[width=3cm]{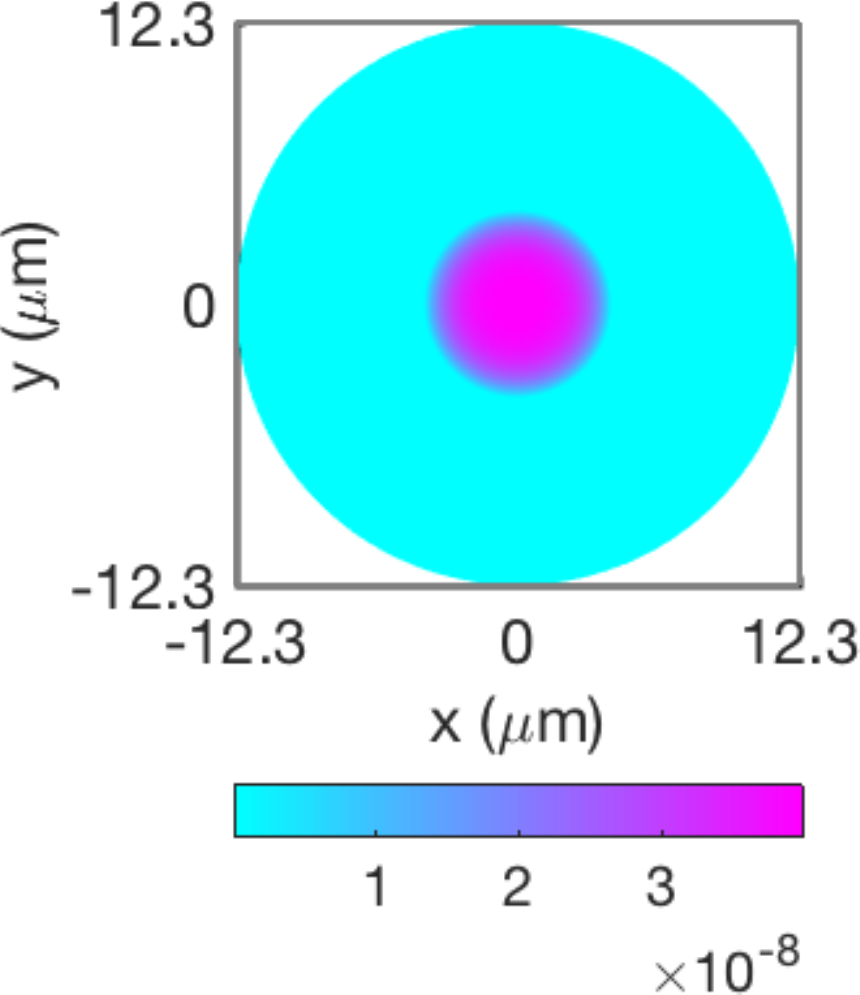}}
\end{minipage}}
\subfigure[3D vector of ${\textbf u}$]{
\begin{minipage}[h]{0.3\linewidth}
\centerline{\includegraphics[scale=0.3]{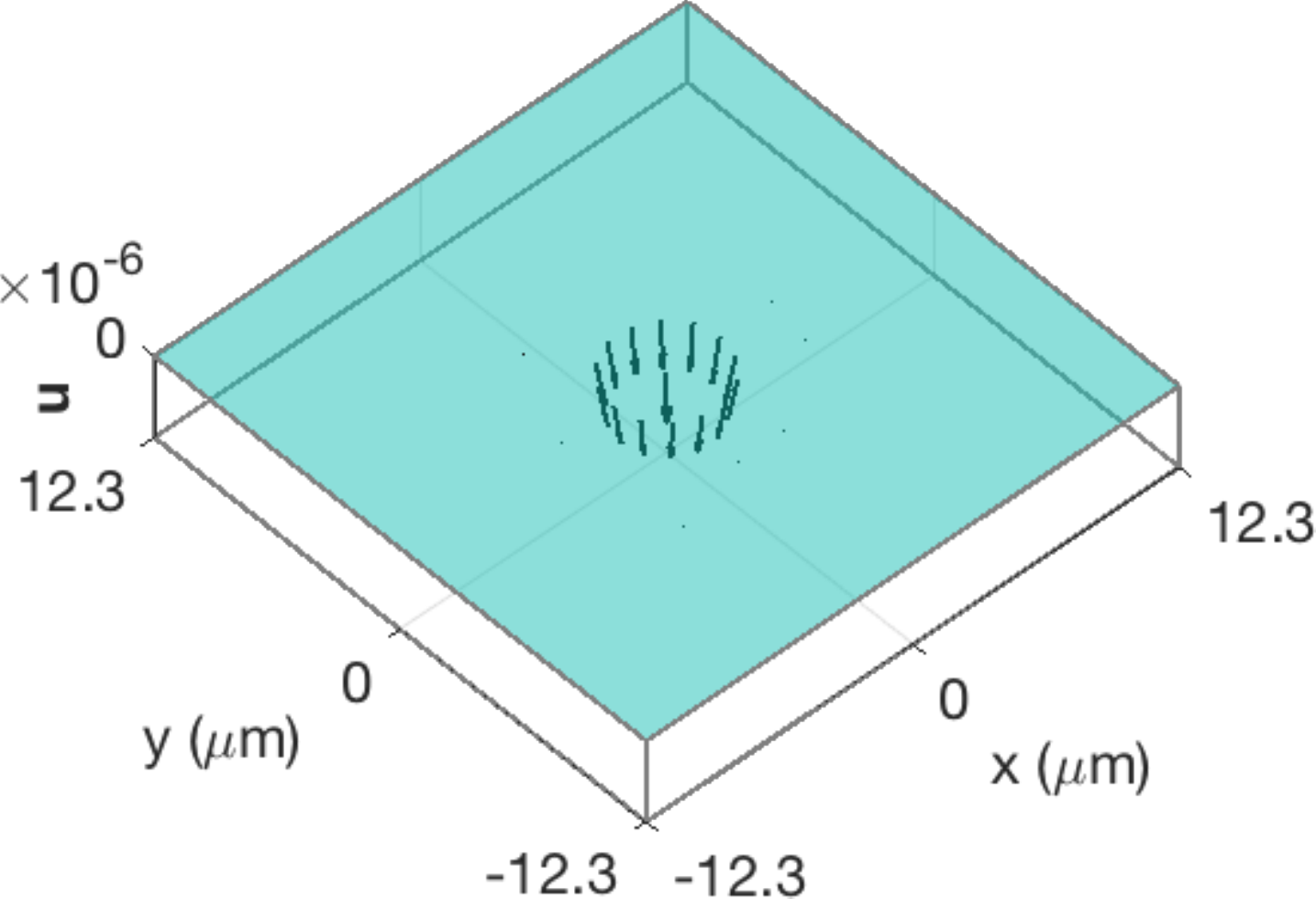}}
\end{minipage}
}
\subfigure[Poynting vector $p_z$]{\begin{minipage}[h]{0.3\linewidth}
\centerline{\includegraphics[width=3cm]{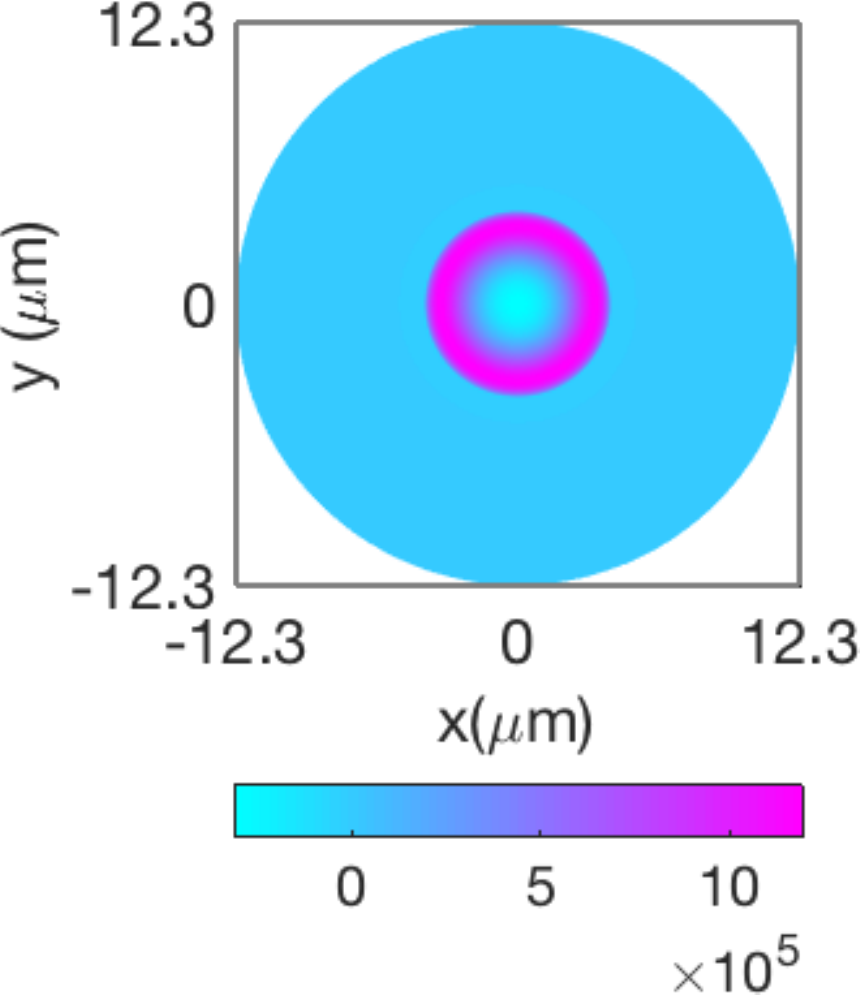}}
\end{minipage}}
\caption{The distributions for ${\textbf u}(x,y)$ for the first eigenmode at low frequency when the waveguide is filled with EMM Core 2 in Figure \ref{dispersion1}. This mode remains the same in the frequency range of $fd/c\in[6.5\times10^{-6},9\times10^{-3}]$. (a) The magnitude distribution of ${\textbf u}(x,y)$. (b) The 3D vector of ${\textbf u}$. (c) The $z$-component of Poynting vector.}
\label{lowfrequency}
\end{figure}

 \item Solid-Fluid Coupling Model\\
In the previous case, the cladding was assumed unbounded, which may not be realistic. Actually, the external medium of the most practical open waveguide problems is fluid  (for example, either air or water). Therefore, here we examine the same size model to verify the fluid-solid coupling system of the open waveguide problems at the frequency of 60 MHz. In this case, the cross section consists of the solid core 3 and the fluid cladding 2. Noting that COMSOL does not provide the ABC  in the modal analysis of the acoustic module. So for comparison, we set the impedance value of the plane wave as an approximation in COMSOL when the outer boundary is far enough. The agreement is  good as illustrated in Table \ref{of_eig_sf}, demonstrating that the proposed SEM is capable of treating the ABC solid-fluid problem. Besides, the relative errors obtained by different orders of SEM confirm the exponential convergence in Fig. \ref{re_sf}. Next, we give a detail discussion about the third mode, whose attenuation constant is almost zero. First, as observed in Fig. \ref{optical_fiber_4}, different from other modes, the propagation of this mode concentrates in the core. The reason for this phenomenon is that the third mode may be caused by the transversal wave, which cannot be transmitted into the fluid region. Moreover, same as the investigation in \cite{1542240}, for an exact integration ( ($N$+1)th-order GLL quadrature in each element) of the second-order geometrical modeling, the errors of mode 3 are straight lines if one groups the even and odd orders separately, and the even and odd orders have different offsets. Therefore the relative error of this mode is reasonable.
 
\begin{table}[th]%
\tbl{$\mathrm{k}_{z} (\times 10^5)$ for Core 3 in water cladding in Figure \ref{optical_fiber_model} obtained by the SEM and FEM.\label{of_eig_sf}}%
{\begin{tabular}  {@{}cccc@{}}
\toprule
$i$& SEM-$\mathrm{k}_{i,z}^5$ & COMSOL-$\mathrm{k}_{i,z}^5$ & SEM-$\bar{\mathrm{k}}_{i,z}^{10}$\\
\hline
1&8.442953-1.01951E-06j&8.442940-1.86170E-05j&8.442953-1.01950E-06j\\
2&7.189511-8.99567E-06j&7.189528-1.14219E-04j&7.189514-8.99557E-06j\\
3&7.079197&7.079197&7.079197\\
4&7.047610-4.23577E-05j&7.048429-9.52793E-04j&7.047614-4.23571E-05j\\
\midrule
Fluid-DOF&5630&50730&88420\\
Solid-DOF&6618&30073&53463\\
\bottomrule
\end{tabular}}
\end{table}

\begin{figure}[th]
\centerline{\includegraphics[scale=0.3]{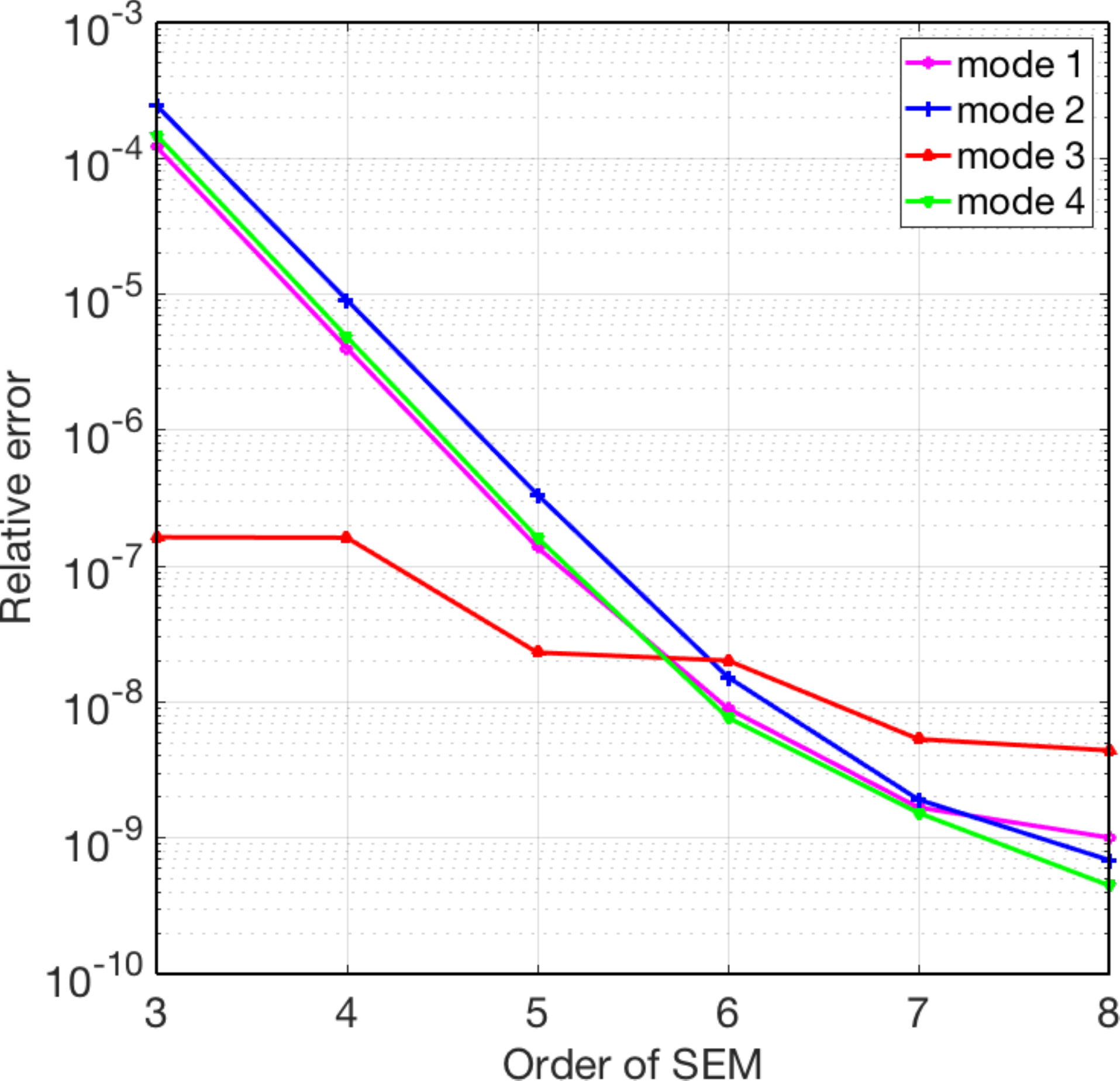}}\vspace*{8pt}
\caption{Relative errors of the first four modes of the solid-fluid open fiber-optics waveguide problem in Figure \ref{optical_fiber_model}. Note that the error curve of the third mode are straight lines if one groups the even and odd orders separately, because the even and odd orders have different offsets.}
\label{re_sf}
\end{figure}
\begin{figure}[th]
\subfigure[Mode for $\mathrm{k}_{1,z}$]{
\begin{minipage}[h]{0.22\linewidth}
\centerline{\includegraphics[width=3cm]{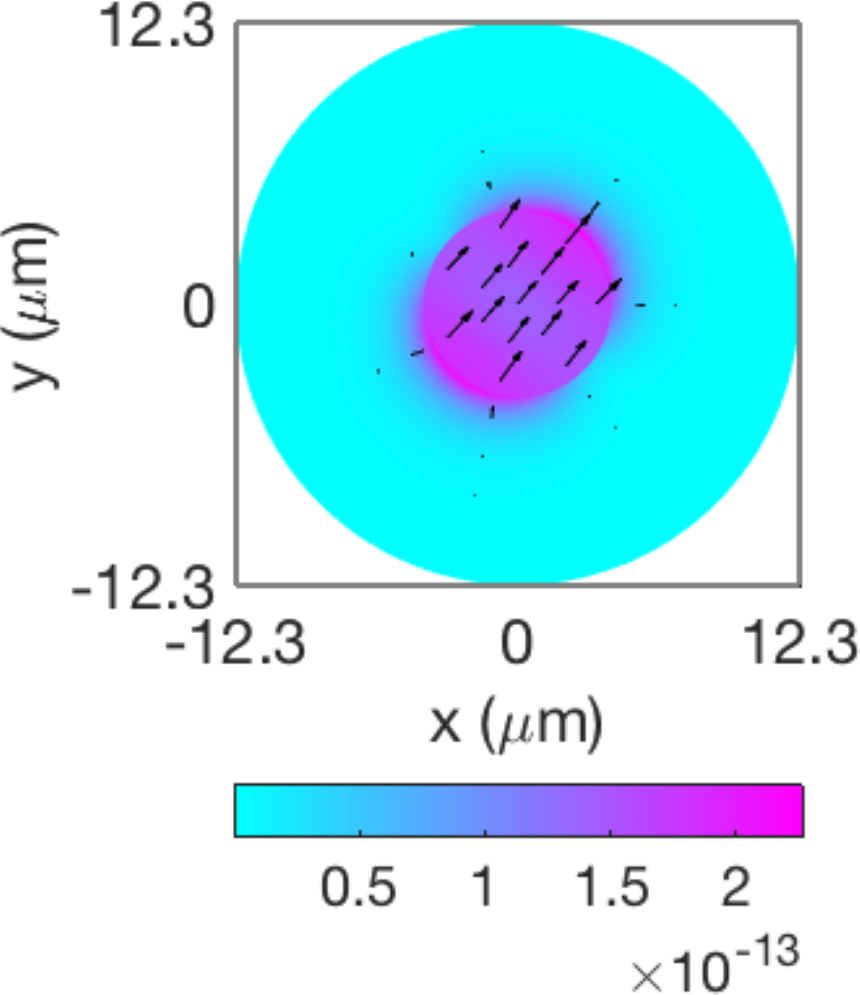}}
\end{minipage}
}
\subfigure[Mode for $\mathrm{k}_{2,z}$]{
\begin{minipage}[h]{0.22\linewidth}
\centerline{\includegraphics[width=3cm]{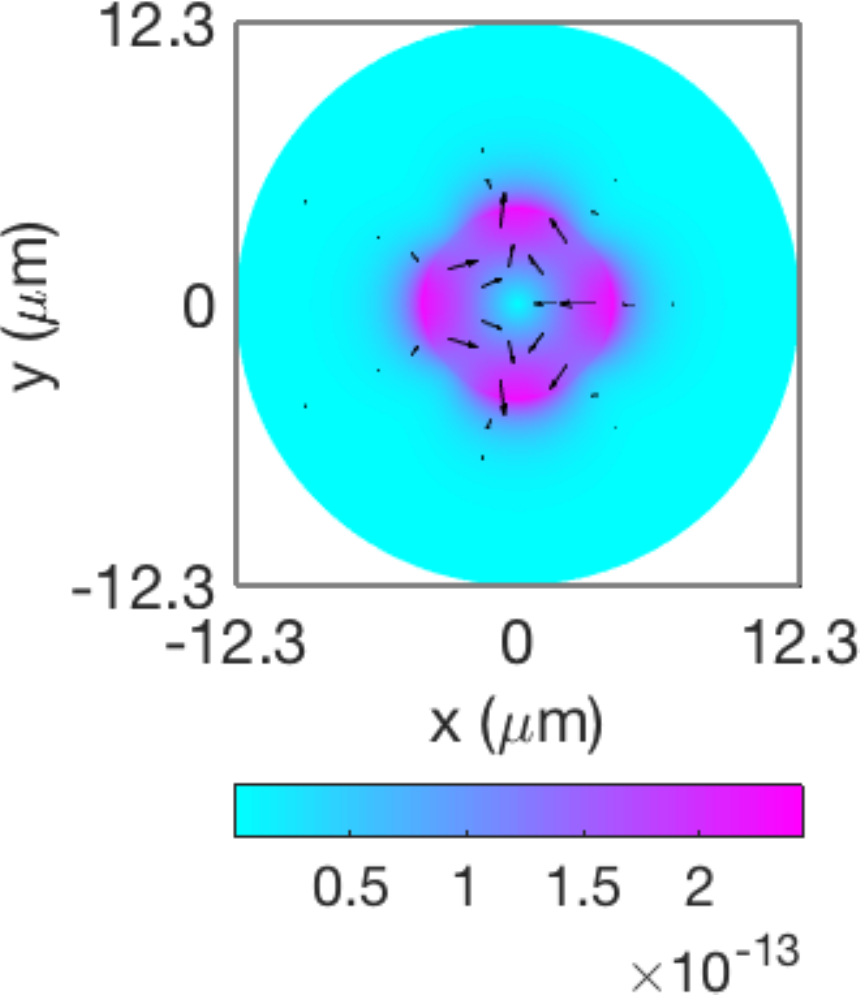}}
\end{minipage}
}
\subfigure[Mode for  $\mathrm{k}_{3,z}$]{
\begin{minipage}[h]{0.22\linewidth}
\centerline{\includegraphics[width=3cm]{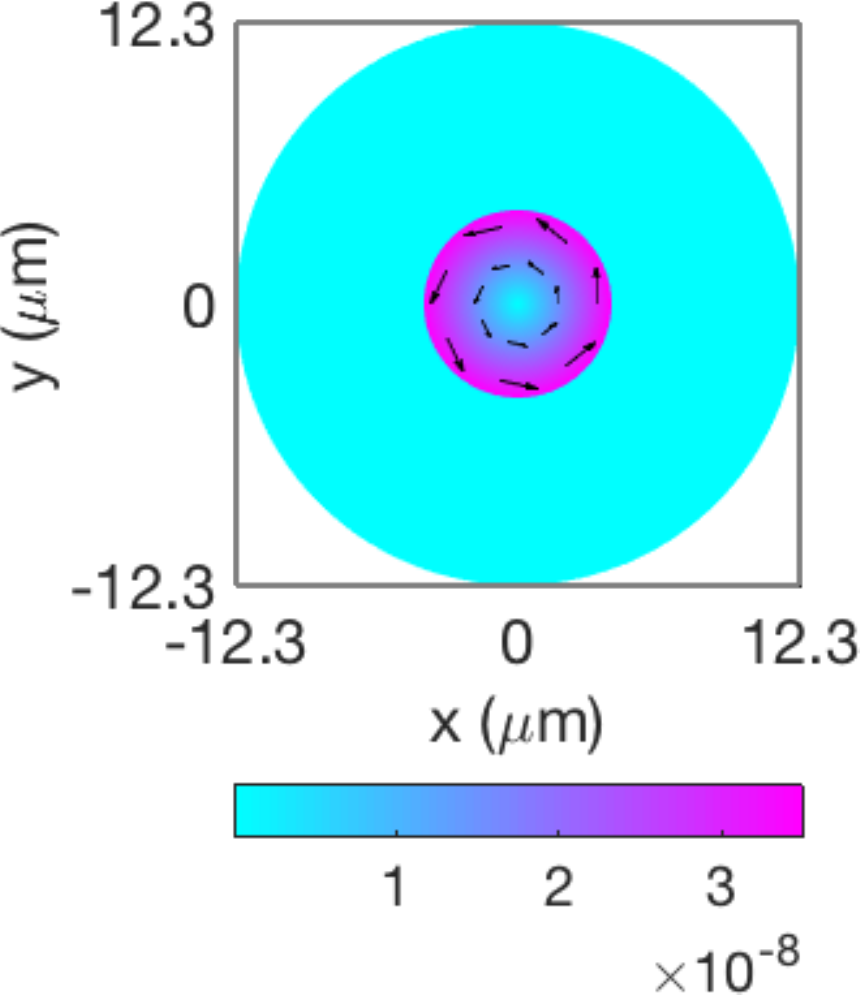}}
\end{minipage}
}
\subfigure[Mode for  $\mathrm{k}_{4,z}$]{
\begin{minipage}[h]{0.2\linewidth}
\centerline{\includegraphics[width=3cm]{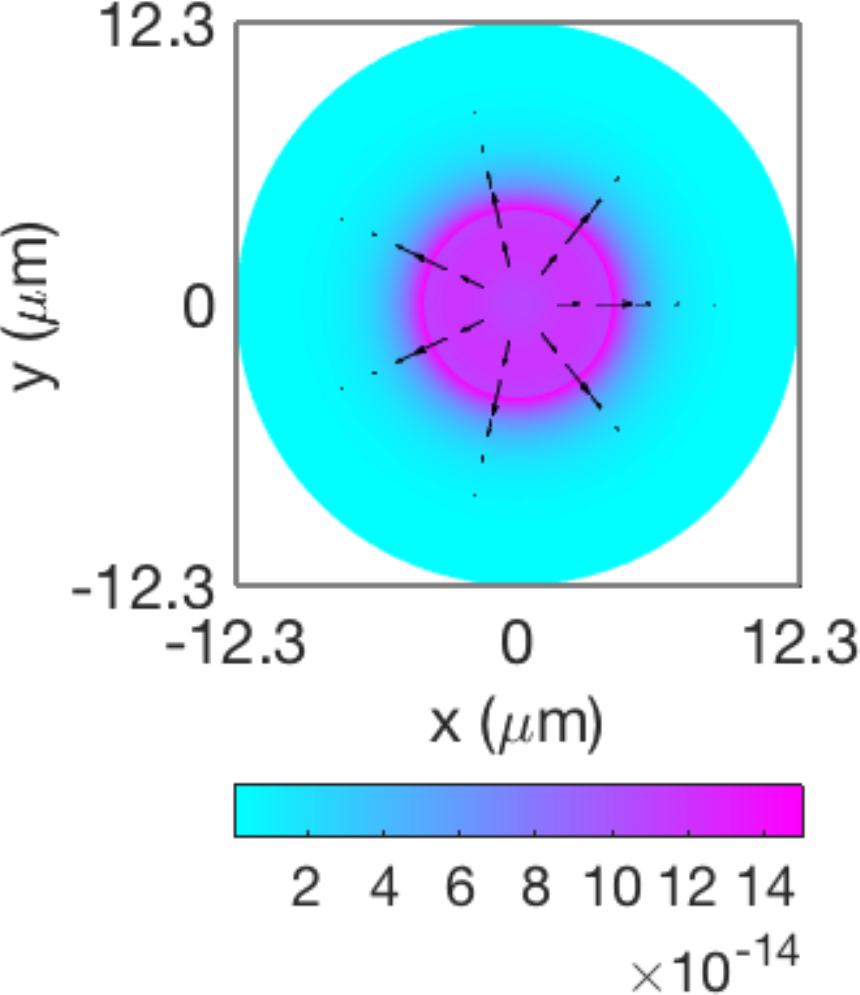}}
\end{minipage}
}\vspace*{8pt}
\caption{Magnitude distributions of ${\textbf u}(x,y)$ for these eignmodes obtained in solid (Core 3) - fluid (Water) open fiber-optics waveguide problem in Figure \ref{optical_fiber_model}. (a)-(d) correspond to the first to the fourth modes. Note that different from other modes, the propagation of the third mode concentrates in the core.}
\label{optical_fiber_4}
\end{figure}
\end{enumerate}
\subsection{EMM with Anisotropic Density}
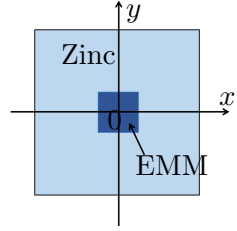
\begin{figure}[th]
\centerline{\begin{overpic}[width=3cm]
                {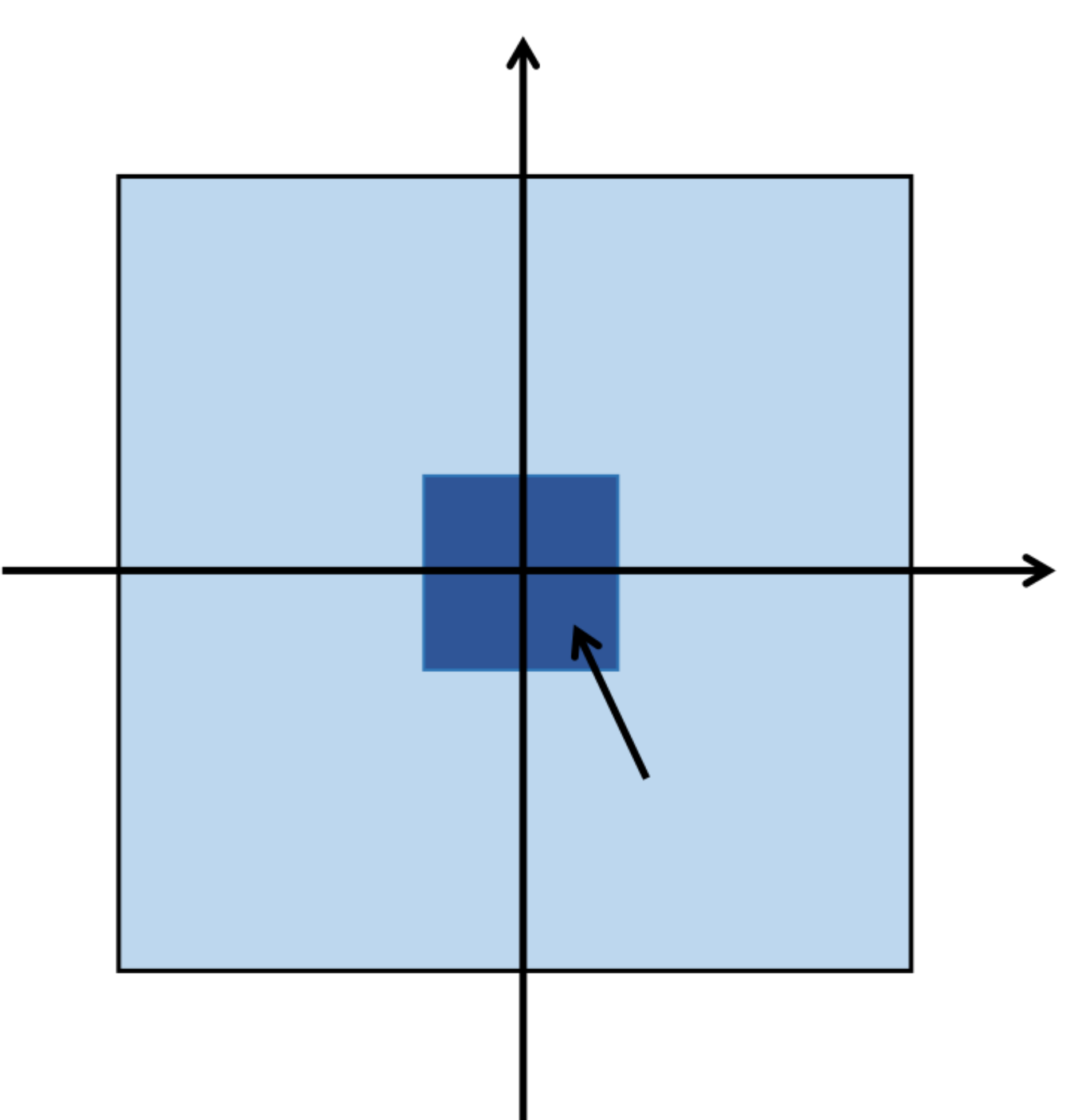}
\put(22,70){Zinc}
\put(54,22){EMM}
\put(50,90){$y$}
\put(42,42){0}
\put(90,52){$x$}
\end{overpic}}\vspace*{8pt}
\caption{The cross section of the anisotropic density waveguide consist of a square EMM core and an unbounded Zinc cladding truncated by a square outer ABC boundary.  }
\label{anisorho_model}
\end{figure}

\begin{figure}[th]
\centerline{
\includegraphics[scale=0.3]{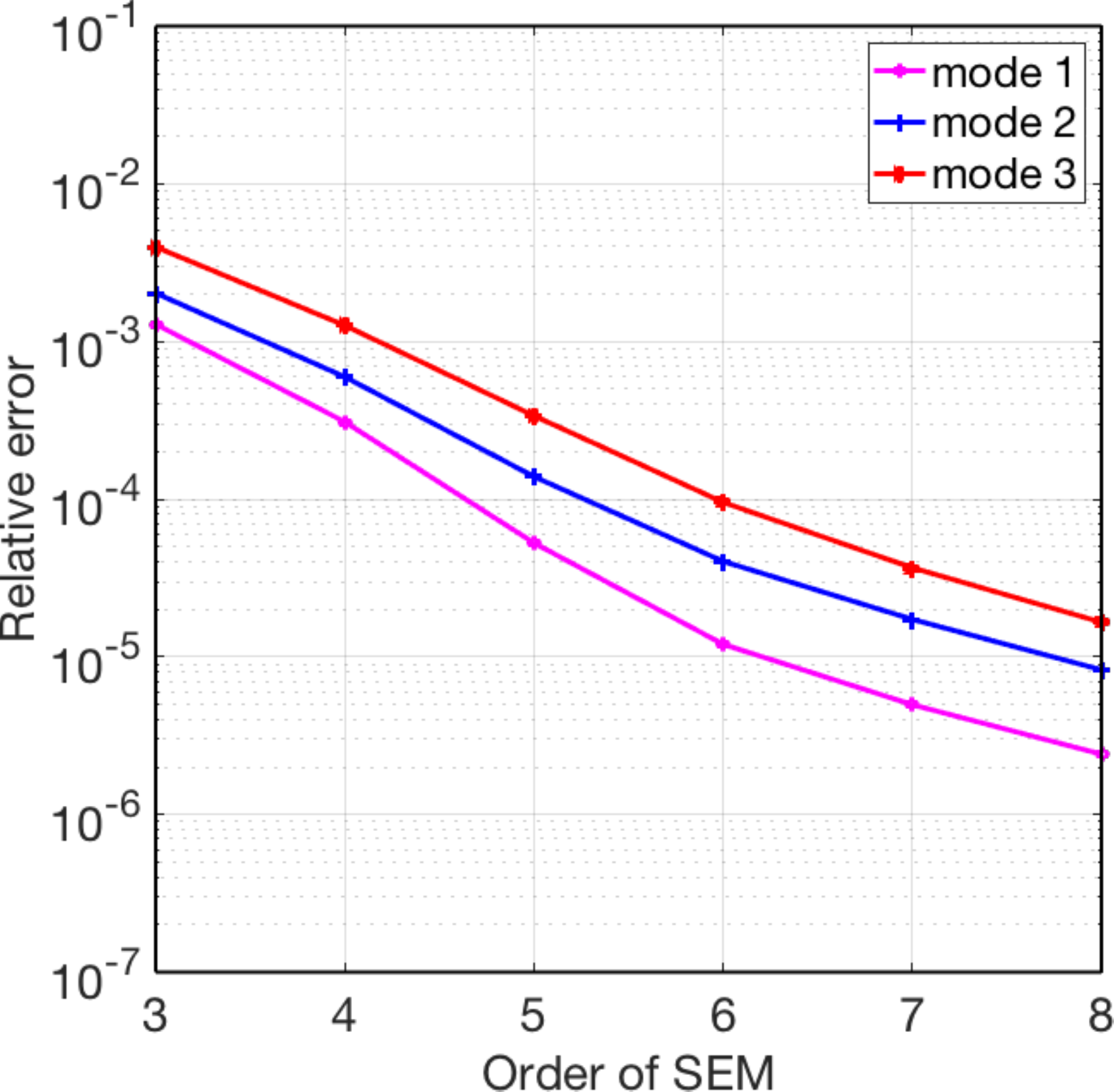}}\vspace*{8pt}
\caption{Relative errors of the first three modes for the anisotropic mass density EMM core waveguide problem in Figure \ref{anisorho_model}.}
\label{anisorho_2}
\end{figure}

\begin{figure}[th]
\subfigure[Mode for $\mathrm{k}_{1,z}$]{
\begin{minipage}[t]{0.3\linewidth}
\centerline{\includegraphics[width=4cm]{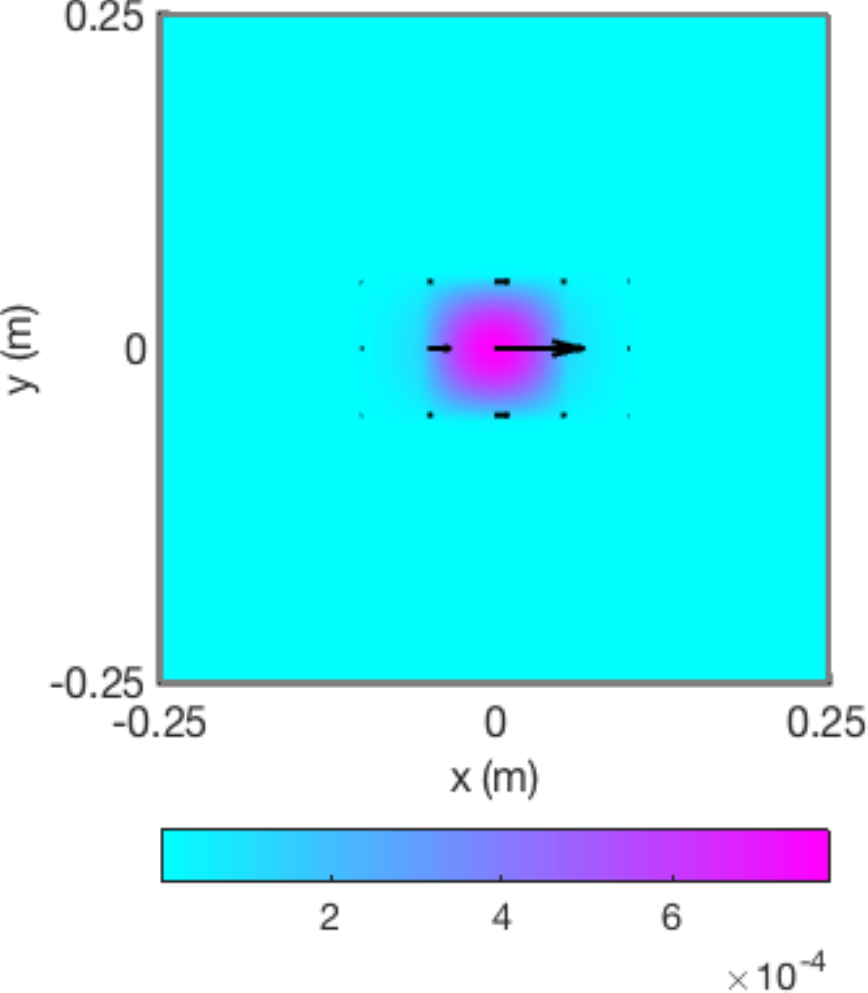}}
\end{minipage}
}
\subfigure[Mode for $\mathrm{k}_{2,z}$]{
\begin{minipage}[t]{0.3\linewidth}
\centerline{\includegraphics[width=4cm]{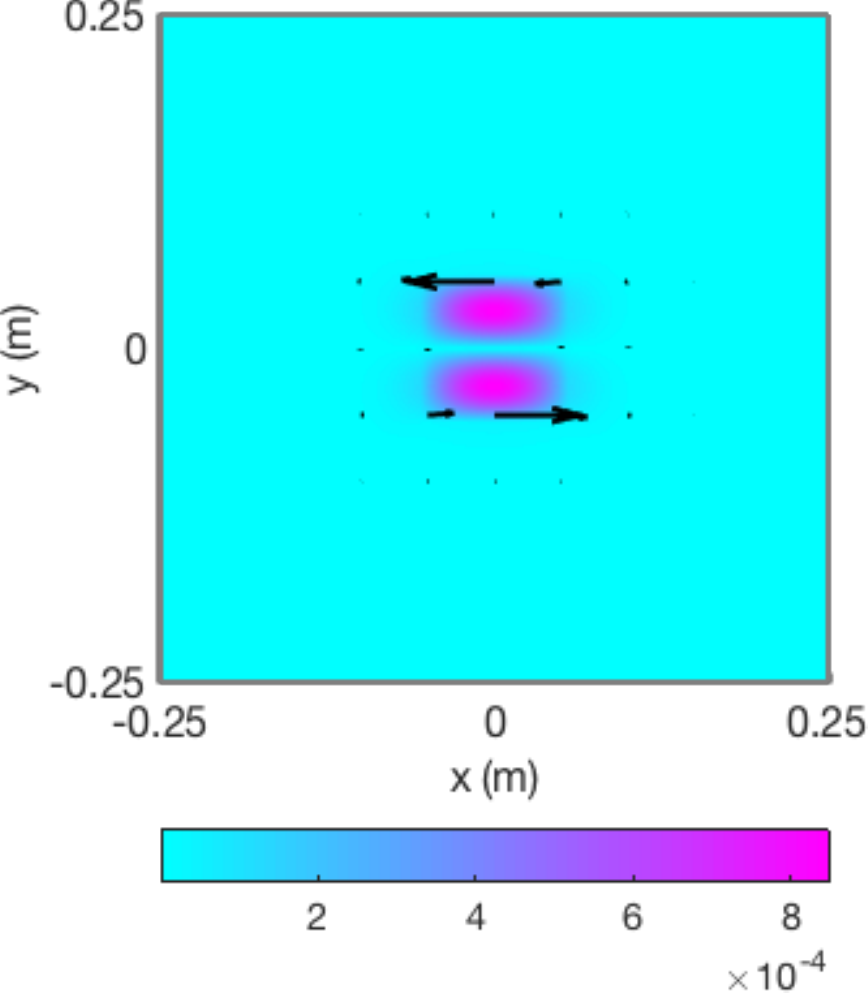}}
\end{minipage}
}
\subfigure[Mode for $\mathrm{k}_{3,z}$]{
\begin{minipage}[t]{0.3\linewidth}
\centerline{\includegraphics[width=4cm]{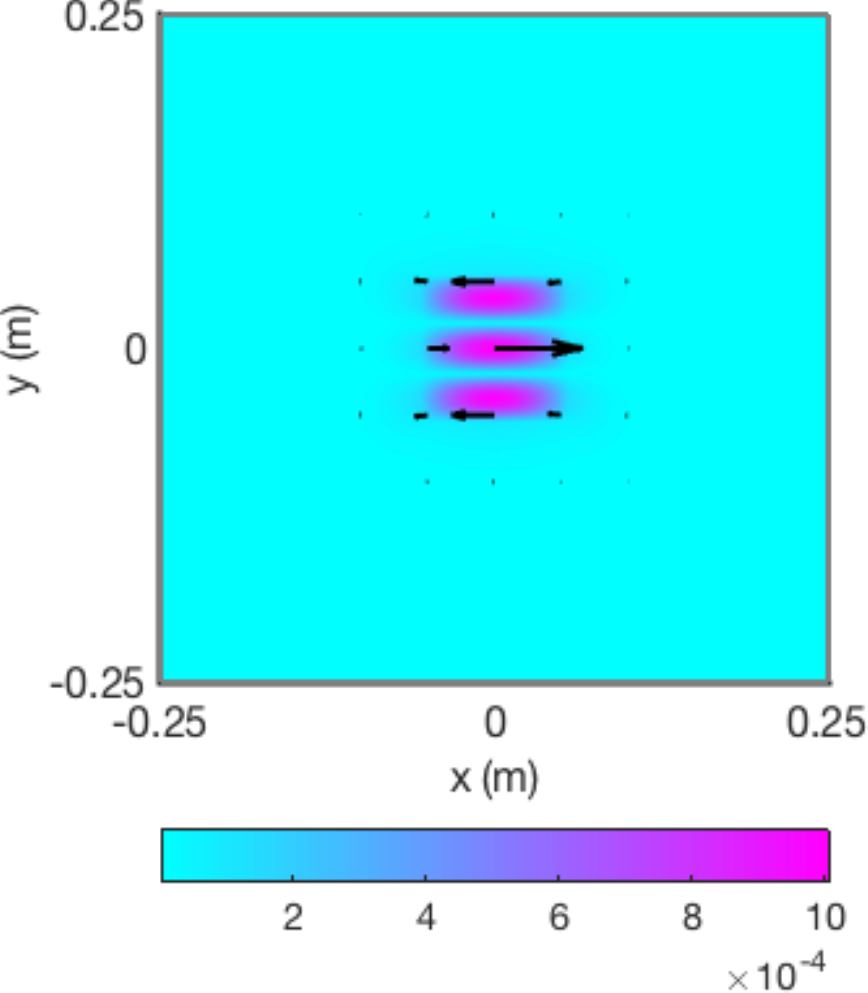}}
\end{minipage}
}\vspace*{8pt}
\caption{Magnitude distributions of $\boldsymbol{\text{u}}$ correspond to $\mathrm{k}_{i,z}$ ($i=1,2,3$) of the anisotropic density EMM core in Figure \ref{anisorho_model}. (a)-(c) correspond to the first to the third modes. All of them are dominated by $u_x$. In contrast to a normal material core, no modes dominated by $u_y$ and $u_z$ are found in this EMM waveguide.}
\label{aniso_3}
\end{figure}

\begin{figure}[th]
\subfigure[Mode for $\mathrm{k}_{1,z}$]
{\begin{minipage}[t]{0.45\linewidth}\centerline{\includegraphics[width=4cm]{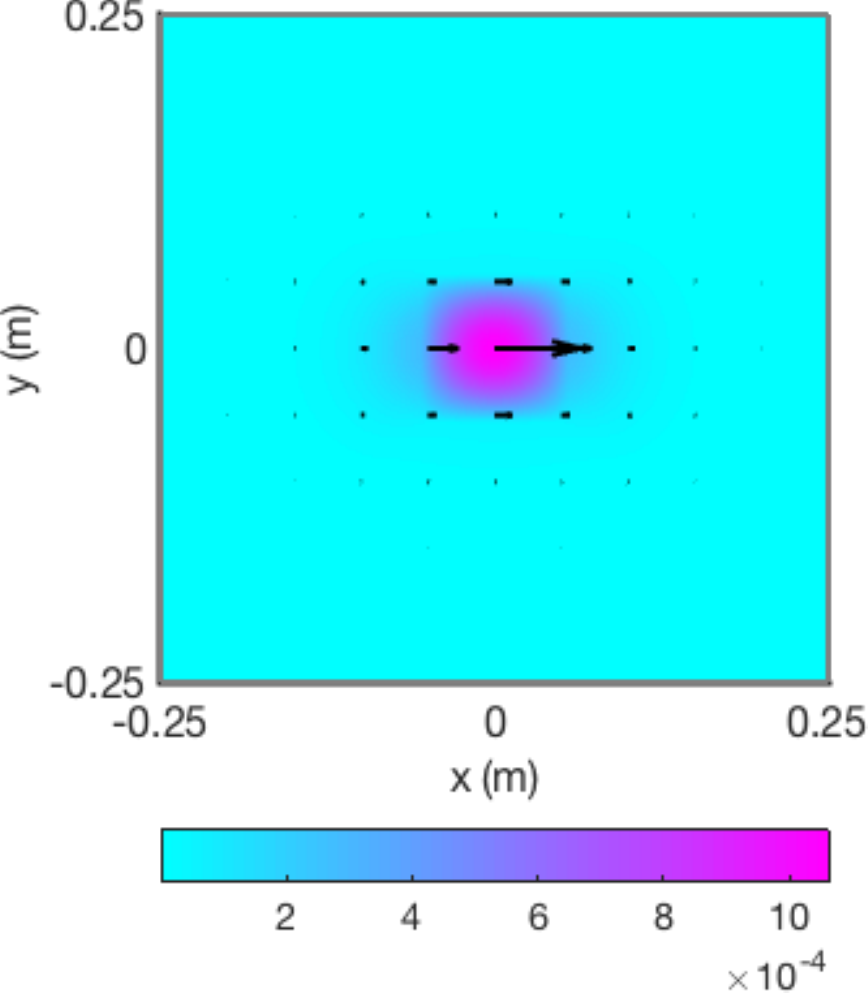}}\end{minipage}}
\subfigure[Mode for $\mathrm{k}_{2,z}$ ]
{\begin{minipage}[t]{0.54\linewidth}\centerline{\includegraphics[width=4cm]{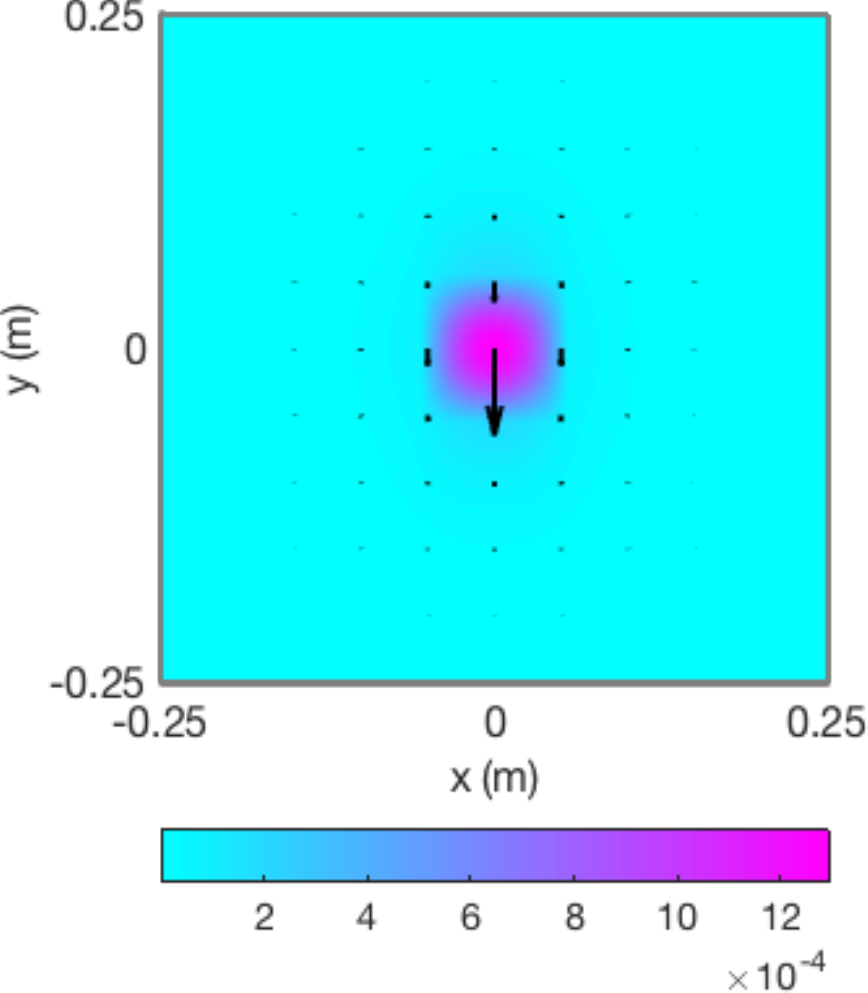}}\end{minipage}}
\subfigure[Mode for $\mathrm{k}_{3,z}$]
{\begin{minipage}[t]{0.45\linewidth}\centerline{\includegraphics[width=4cm]{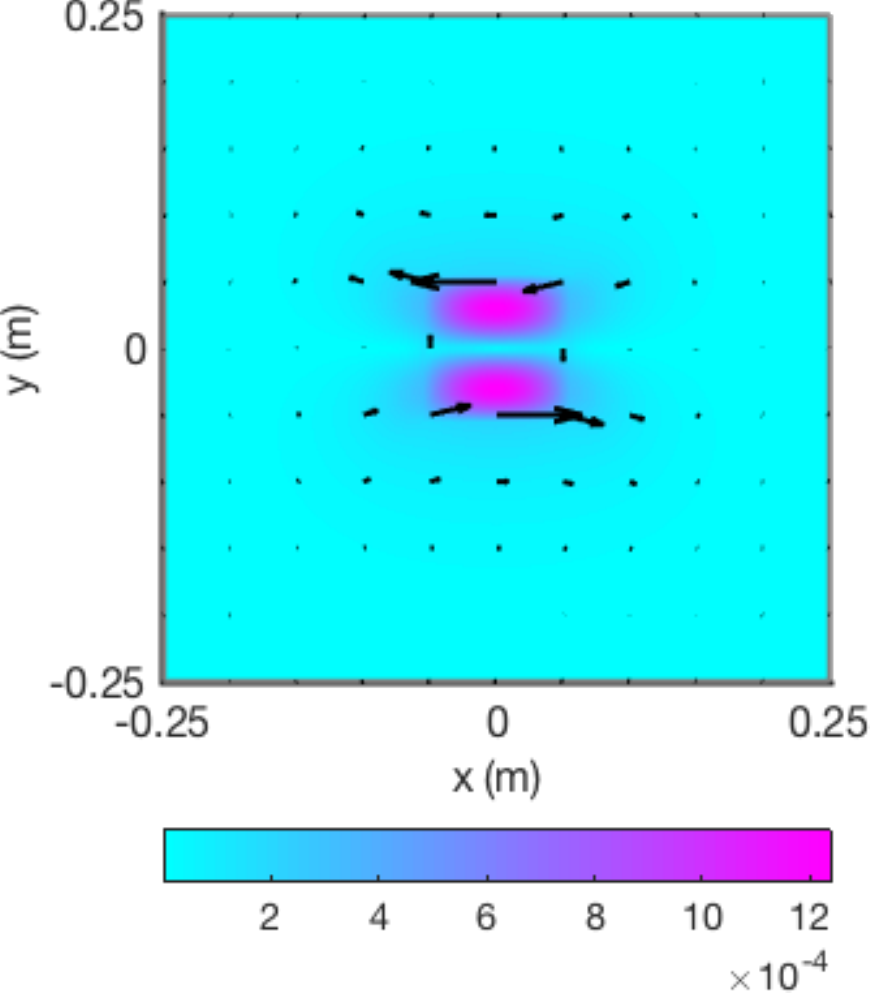}}\end{minipage}}
\subfigure[Mode for $\mathrm{k}_{4,z}$]
{\begin{minipage}[t]{0.53\linewidth}\centerline{\includegraphics[width=4cm]{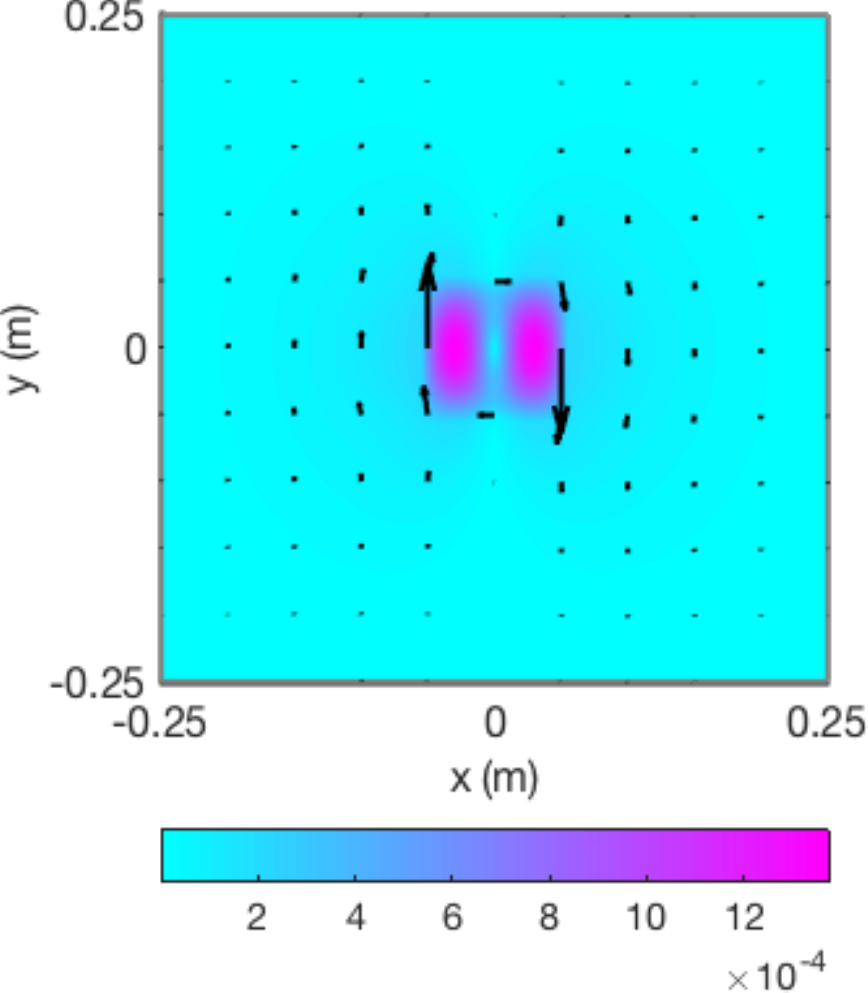}}\end{minipage}}
\caption{Magnitude distributions of $\boldsymbol{\text{u}}$ for eignmodes $\mathrm{k}_{i,z}$ ($i=1,\cdot,4$) obtained in an open elastic waveguide problem with a normal isotropic density core with the EMM core in Figure \ref{anisorho_model} replaced by an isotropic mass density
$\rho = 3772.5$ kg/m$^3$. (a)-(d) correspond to the first to the fourth modes.  Note that the 2nd, 4th modes dominated by $u_y$ and $u_z$ respectively are absent in the EMM waveguide in Fig. \ref{aniso_3}.}
\label{aniso_4}
\end{figure}

In addition to the metamaterials with negative index discussed above, the metamaterials with anisotropic mass density have attracted more and more attention recently.  Because the equivalent model with effective anisotropic mass density can describe the dynamic behavior of the original lattice system in all directions. Hence, we conduct one numerical experiment on one anisotropic density core that cannot be simulated by some traditional numerical methods. Besides, in this section, the formulation of phase velocity obtained by elastic tensor $C$ and isotropic density $\rho$ in literature \cite{Thomas1997Elastic} is extended to one suitable for anisotropic density $\bar{\bar{\rho}}=(\rho_{ij})_{3\times3}$. Note that the explanations of the symbols are referred to reference \cite{Thomas1997Elastic}. Starting with time domain governing equation 
\begin{equation}\label{time gover}\boldsymbol{\rho}\cdot\frac{\partial^2\boldsymbol{\text{u}}}{\partial t^2}-\nabla\cdot\boldsymbol{\tau}=0\end{equation}  Multiplying both sides by the inverse of $\bar{\bar{\rho}}$, the scalar expressions for a homogeneous medium are arrived at
\begin{equation}\label{time gover2}
\frac{\partial^2u_i}{\partial t^2}=(\boldsymbol{\rho}^{-1})_{im}\cdot c_{mjkl}\frac{\partial^2u_k}{\partial x_l\partial x_j}
\end{equation}
After denoting $\Gamma_{ik}=c_{ijkl}n_jn_l$, where $n_j$ is the component of the unit propagation vector ${\hat{\textbf n}}$ and multiplying both sides by $p_l$, the components of the unit polarization vector colinear with the displacement. The final eigenvalue formulation is obtained
  \begin{equation}\label{time gover3}
[(\boldsymbol{\rho}^{-1})_{im}\cdot\Gamma_{ml}-v^2\delta_{il}]p_l=0
\end{equation}
where $v^2$ is the eigenvalue. The cross section centered at (0,0) m is shown in Fig. \ref{anisorho_model}, the width of the square cladding and the square core is 0.5 m and 0.11 m, respectively. The cladding is zinc with isotropic material parameters $\{v_p,v_s\}=\{4820.7,2361.6\}$ m/s and the mass density is 7100 kg/m$^3$. The core is an anisotropic EMM with the effective elastic coefficients $\text{C}_{11}=36.63 ~\text{GPa}$, $\text{C}_{12}=5.57 ~\text{GPa}$, $\text{C}_{13}=13.53 ~\text{GPa}$, $\text{C}_{22}=18.83 ~\text{GPa}$, $\text{C}_{23}=7.84 ~\text{GPa}$, $\text{C}_{33}=48.38~ \text{GPa}$, $\text{C}_{44}=12.41 ~\text{GPa}$, $\text{C}_{55}=6.69 ~\text{GPa}$, $\text{C}_{66}=2.272 ~\text{GPa}$. The frequency we choose is $f=16$ kHz and corresponding effective anisotropic mass density represent  $\rho_{\text{EMM}}=\text{diag}\{6277,3168,2700\}$ $\text{kg/m}^3$ according to \cite{ZhuMicrostructural}. The anisotropic mass density is frequency-dependent and caused by the different locally resonant frequencies along different directions in the microstructure design, depending on the inverse proportional function $\rho_{\text{eff},i}=a+b/(\omega_i^2-\omega^2)$ \cite{ZhuMicrostructural,10.1007/978-90-481-9893-1_14}, where $a,b$ are the positive constants given by the detailed model parameters and $\omega_i$ is the locally resonance frequency along the $i$ direction ($i=x,y,z$). Note that $\omega_1$ is the smallest, the $\rho_{\text{EMM}}$ is certainly produced by the frequency below and close to the $\omega_1$, leading to the resonance phenomena dominated by the $u_x$. Besides, the velocity of EMM along ${\hat{\textbf n}}=(0,0,1)$ calculated through equation (\ref{time gover}) is $\{v_p,v_{s1},v_{s2}\}=\{4233,1979.2,1032.4\}$ m/s, smaller than the cladding. Therefore, the ABC is used to truncate the cladding. First, the good agreement between the 5th-order SEM numerical results and the 10th-order results of the extremely fine mesh is demonstrated in Table \ref{aniso}. Then, the relative errors of the first three modes plotted in Fig. \ref{anisorho_2} indicate the exponential convergence. Moreover, the magnitude distributions of ${\textbf u}$ are plotted in Fig. \ref{aniso_3} and all of them in the $xy$ plane are along the $x$ direction and dominated by the $x$-component of the ${\textbf u}$. On the other hand, we conduct another experiment on an normal anisotropic elastic core with the isotropic mass density $\rho=3772.5$ $\text{kg/m}^3$ (the geometric average$ \sqrt[3]{\prod_{i=1}^{3}\rho_{ii}}$) and the same elastic coefficients $C$.  The magnitude distributions of ${\textbf u}$ in this waveguide with normal materials are plotted in Fig. \ref{aniso_4} and they are dominated by the components along three principal axis respectively. In contrast to the configuration with an EMM core,  it can be found that the propagation mode dominated by $u_y$, $u_z$ shown in Fig. \ref{aniso_4} (c),(d) cannot be obtained in the example with the anisotropic mass density core. The phenomena are due to the difference between the EMM core and the normal core in view of the locally resonance frequencies in each principal axis, which is caused by the different mass density tensors. In conclusion, the above explains the phenomena caused by the use of the EMM core with the anisotropic mass density and demonstrates the rationality of our results.

\begin{table}[th]%
\tbl{$\mathrm{k}_{i,z}^\mathrm{n}$ of the anisotropic mass density EMM waveguide problem in Figure \ref{anisorho_model}.\label{aniso}}%
{\begin{tabular}  {@{}ccc@{}}
\toprule
 $i$ & SEM~-~$\mathrm{k}_{i,z}^5$ & SEM~-~$\bar{\mathrm{k}}_{i,z}^{10}$\\
\midrule
1 &79.78808-6.2E-10j&79.78866-6.2E-10j\\
2 &73.91172-1.1E-09j&73.91355-1.1E-09j\\
3 &63.19895-1.3E-07j&63.20232-1.3E-07j\\
\midrule
Mesh&289&625\\
DOF&22188&189003\\
\bottomrule
\end{tabular}}
\end{table}
 

\section{Conclusions}
This paper presents a SEM solver for the general EMM waveguide problems with negative index and anisotropic mass density as well as normal materials. The solver can treat inhomogeneous and anisotropic solids, but also include the fluid-solid coupling. Meanwhile, the discussions about four boundary conditions (the hard BC, the soft BC, the BPBC, the ABC) are provided. Both excellent agreement between results and those from the commercial FEM solver COMSOL and less computational costs are demonstrated in the numerical validations.  Moreover, some interesting phenomena brought by the application of the EMM can be observed in the numerical experiments, for instance, unusual modes generated by the negative refractive index or common modes eliminated by the anisotropic mass density.

\section*{Acknowledge}
This work was supported by the National Key Research and Development Program of
China [grant numbers 2018YFC0603503]; and the China Postdoctoral
Science Foundation [grant numbers 2019M662244].
\bibliographystyle{ws-jtca}
\bibliography{REF}

\end{document}